\documentclass[a4paper,aps,pre,10pt,floatfix,twocolumn,noshowpacs,superscriptaddress,longbibliography]{revtex4-1}
\usepackage{graphicx}
\usepackage[pdftex]{hyperref}
\usepackage{amssymb}
\usepackage{amsmath}
\usepackage{amsfonts}
\usepackage{pdfpages}
\begin{document}
%
%
%
%
\title{The geometric nature of weights in real complex networks}
\author{Antoine \surname{Allard}}
\affiliation{Departament de F\'isica de la Mat\`eria Condensada, Universitat de Barcelona, Mart\'i i Franqu\`es 1, E-08028 Barcelona, Spain}
\affiliation{Universitat de Barcelona Institute of Complex Systems (UBICS), Universitat de Barcelona, Barcelona, Spain}

\author{M. \'Angeles \surname{Serrano}}
\affiliation{Departament de F\'isica de la Mat\`eria Condensada, Universitat de Barcelona, Mart\'i i Franqu\`es 1, E-08028 Barcelona, Spain}
\affiliation{Universitat de Barcelona Institute of Complex Systems (UBICS), Universitat de Barcelona, Barcelona, Spain}
\affiliation{Instituci\'o Catalana de Recerca i Estudis Avan\c{c}ats (ICREA), Passeig Llu\'is Companys 23, E-08010 Barcelona, Spain}

\author{Guillermo \surname{Garc\'ia-P\'erez}}
\affiliation{Departament de F\'isica de la Mat\`eria Condensada, Universitat de Barcelona, Mart\'i i Franqu\`es 1, E-08028 Barcelona, Spain}
\affiliation{Universitat de Barcelona Institute of Complex Systems (UBICS), Universitat de Barcelona, Barcelona, Spain}

\author{Mari\'an \surname{Bogu\~n\'a}}
\thanks{Corresponding author: marian.boguna@ub.edu}
\affiliation{Departament de F\'isica de la Mat\`eria Condensada, Universitat de Barcelona, Mart\'i i Franqu\`es 1, E-08028 Barcelona, Spain}
\affiliation{Universitat de Barcelona Institute of Complex Systems (UBICS), Universitat de Barcelona, Barcelona, Spain}
\date{\today}
\begin{abstract}
  The topology of many real complex networks has been conjectured to be embedded in hidden metric spaces, where distances between nodes encode their likelihood of being connected. Besides of providing a natural geometrical interpretation of their complex topologies, this hypothesis yields the recipe for sustainable Internet's routing protocols, sheds light on the hierarchical organization of biochemical pathways in cells, and allows for a rich characterization of the evolution of international trade. We present empirical evidence that this geometric interpretation also applies to the weighted organisation of real complex networks. We introduce a very general and versatile model and use it to quantify the level of coupling between their topology, their weights, and an underlying metric space. Our model accurately reproduces both their topology and their weights, and our results suggest that the formation of connections and the assignment of their magnitude are ruled by different processes.
\end{abstract}
\maketitle
%
%
%
%
%
\section*{Introduction}
%

Most of the complexity of networks is encoded into the intricate topology of the interactions among their components and into the layout of the intensities associated to such interactions (i.e., the weights). Interestingly, weights are coupled in a non-trivial way to the binary network topology, playing a central role in their structural organisation, function, and dynamics~\cite{Barrat2004a}. For instance, the quantification of the rich-club effect in real weighted networks, in sharp contrast to results in unweighted representations, unveils the formation of alliances in multipolarised environments or the lack of cohesion even in the presence of rich-club ordering~\cite{Serrano:2008b}. Similarly, the propagation of emergent diseases in the international airports network is intimately linked to the number of passengers flying from one airport to the other~\cite{Brockmann2013}. A shift towards a paradigm of weighted networks is therefore in order to fully understand the behaviour and evolution of complex networks. However, advances in this area have been limited by the extreme heterogeneity and fluctuations that typically characterise the distribution of weights. 

Meanwhile, complex networks~\cite{NEWMAN2010,Cohen10_ComplexNetworks} have been conjectured to be embedded in hidden metric spaces, in which distances among nodes encode a balance between their similarity and popularity and, thus, determine their likelihood of being connected~\cite{Serrano08_PhysRevLett}. This hypothesis, combined with a suitable underlying space, has offered a geometric interpretation of the complex topologies observed in real networks, including scale-free degree distributions, the small-world effect, strong clustering, community structure, and self-similarity. A metric space underlying complex networks can also explain their efficient inter-node communication without knowledge of the complete structure~\cite{Boguna2009,Boguna2009c}. Moreover, it has been shown that for networks whose degree distribution is scale-free, the natural geometry of their underlying metric space is hyperbolic~\cite{Krioukov2010,gugelmann2012,bode2013,candellero2014,friedrich2015,Aste2005}. All these results have then been used to propose geometric models for real growing networks that reproduce their evolution and in which preferential attachment emerges from local optimisation principles~\cite{Papadopoulos2012,Gulyas2015}. Finally, mapping real complex networks into a hidden metric space has yielded a sustainable solution to the scaling limitations of the Internet~\cite{Boguna10_NatCommun}, has shed light on the hierarchical organisation of biochemical pathways in cells~\cite{Serrano2012}, and has allowed a rich characterization of the evolution of international trade over fourteen decades \cite{Garcia-Perez2016}.

In real weighted networks, weights are coupled to the binary topology in a non-trivial way. This is manifested, for instance, in a non-linear relation between the strength of a node $s$ (the sum of the total weight attached to it) and its degree $k$ of the form $s \sim k^{\eta}$~\cite{Barrat2004a,Popovic2012,Barthelemy11_PhysRep}. However, the relation between the layout of weights and the geometry underlying the network is unclear. The reason being that, even if the existence of a link depends on the metric distance between the nodes, there is no reason, a priori, to expect that the same distance will affect its weight. For instance, in the airports network, the decision to set up a link between two cities depends on the airline companies operating at the two airports, a process affected by geopolitic and economic costs and by the expected flow of passengers that would eventually compensate such costs. However, once the connection is established, its weight is determined by the aggregation of the individual decisions of people using it, a process that may be affected by a different cost function.

In this paper, we present empirical evidence on the metric nature of weights in real biological, economic and transportation networks (see Methods for a description of the datasets), which suggests that the hidden/latent geometry paradigm can be extended to weighted complex networks. We then propose a general class of weighted networks embedded in hidden metric spaces that accurately reproduces many properties observed in real weighted networks. This model has the critical ability to fix the degree-strength distribution independently of the coupling of the topology and weighted organisation with the metric space. It is therefore possible to isolate, and thus directly study, the effect of the coupling between the metric space and the weighted organisation of real weighted networks. In fact, our results unveil that in some systems these couplings are uncorrelated, which in turn suggests that the formation of connections and the assignment of their magnitude might be ruled by different processes. Our empirical findings, combined with our new class of models, open the path towards the use of information encoded in the weights of the links to find more accurate embeddings of real networks, which in turn will improve the detection of communities, the prediction of missing links and provide estimates for the weights of such missing links.
%
%
%
%
%
\section*{Results}
%
%
%
\subsection*{Interplay between weights and triangles in real networks}
\begin{figure}[t] 
\includegraphics[width = 0.9\linewidth]{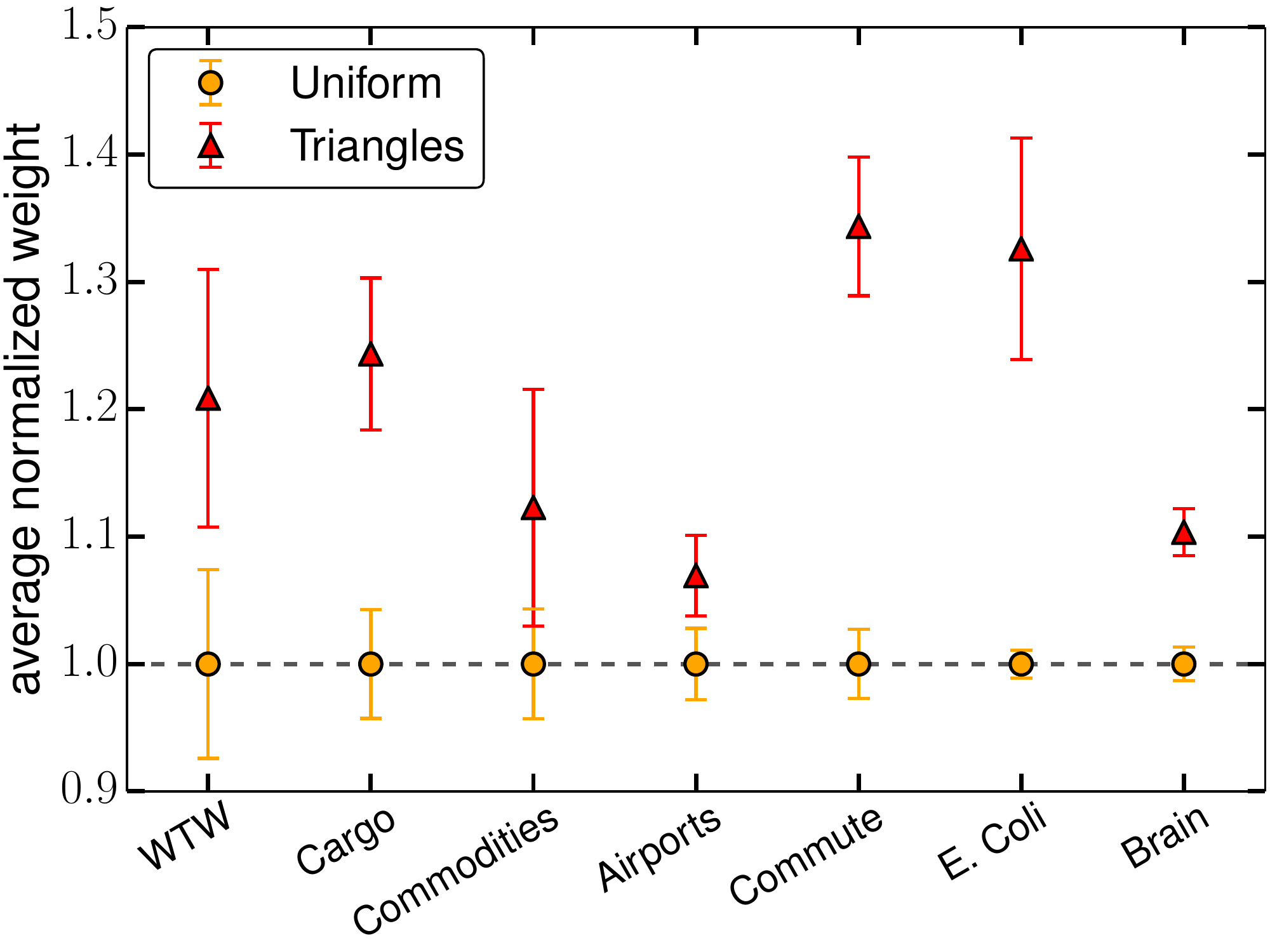}
\caption{\textbf{Geometric nature of weights.} Comparison of the average normalised weights in the network (yellow circles) with the one measured by sampling links over triangles (red triangles) for the empirical datasets analysed. The error bars correspond to an estimate of the standard deviation of the average value due to the finite size of the samples and are computed as $\sqrt{\mathrm{Var}[\omega^\mathrm{norm}]/L}$, where $\mathrm{Var}[\omega^\mathrm{norm}]$ is the variance of the normalised weights sampled uniformly or via the triangles, and $L$ is the number of links.}
\label{fig:1}
\end{figure}

%
Clustering, as a reflection of the triangle inequality, is the key topological property coupling the bare topology of a complex system and its effective underlying metric space~\cite{Serrano08_PhysRevLett}. In this context, the triangle inequality stipulates that if nodes A and B are close, and nodes A and C are also close, we expect nodes B and C to be close as well; triangles are therefore more likely to exist between nodes that are nearby. Consequently, we expect that if the weights of connections depend on the distance between the connected nodes in the underlying metric space, they should be quantitatively different depending on the clustering properties of the connections. However, weights and clustering are known to be strongly influenced by the degrees of endpoint nodes \cite{Barrat2004a,Popovic2012,Pajevic2012}, which prevents from a direct detection of the metric properties of weights due to the typical heterogeneity in the degrees of nodes in real networks. Thus, to compare links on an equal footing, we define the normalised weight of an existing link connecting nodes $i$ and $j$ as $\omega_{ij}^\mathrm{norm}=\omega_{ij}/\bar{\omega}(k_ik_j)$, where $\bar{\omega}(kk')$ is the average weight of links as a function of the product of degrees of their endpoint nodes. By doing so, we decouple the weights and the topology, leaving the normalised weights seemingly randomly fluctuating around 1 (see uniform sampling on Fig.~\ref{fig:1}).

Figure~\ref{fig:1} shows, however, that these fluctuations are not uniform as links involved in triangles tend to have larger normalised weights than the average link. Indeed, in some cases the difference can reach more than $30$\%. Sampling links over triangles is equivalent to sampling links proportionally to their multiplicity $m$ (i.e., the number of triangles to which a link participates). Therefore, the results in Fig.~\ref{fig:1} indicate that $\omega^\mathrm{norm}$ and $m$ are positively correlated variables, as corroborated by their Pearson correlation coefficient (see Supplementary~Table~1). In Ref.~\cite{Pajevic2012}, the authors also found local correlations between the multiplicity of links and the weights for different real networks. Note however that in that study weights were not normalised to discount the effects of the heterogeneity in the degrees of the nodes, so that the detected weighted organisation can not be taken as a signature of underlying metric properties.

Since triangles are a reflection of the triangle inequality in the underlying metric space, we expect nodes forming triangles to be close to one another. Thus, the higher average normalised weight observed on triangles strongly suggests a metric nature of weights, which is not a trivial consequence of the relation between weights and topology. This leads us to formulate the hypothesis that the same underlying metric space ruling the network topology---inducing the existence of strong clustering as a reflection of the triangle inequality in the underlying geometry---is also inducing the observed correlation between $\omega^\mathrm{norm}$ and $m$. To prove this, we develop a realistic model of geometric weighted random networks, which allows us to estimate the coupling between weights and geometry in real networks.
%
%
%
%
%
%
%
%
\subsection*{A geometric model of weighted networks}
%
\begin{table*}[t]
  \begin{ruledtabular}
    \begin{tabular}{llllllllll|l|l|l|l|}
      \textbf{Name}  & \textbf{Type}           & \textbf{Nodes}   & \textbf{Weights}   & $\gamma$ & $\eta$ & $\langle k \rangle$ & $N$  \\ \hline
      World Trade    & Economic                & Countries        & US dollars         & 2.42     & 1.6    & 5.8                & 189  \\ 
      Cargo ships    & Transportation          & Ports            & Shipping journeys  & 2.03     & 1.05   & 10.4                & 834  \\
      US Commodities & Economic                & Economic sectors & US dollars         & 2.46     & 1.22   & 5.8                 & 376  \\
      US Airports    & Transportation          & Airports         & Passengers         & 1.76     & 1.7    & 8.6                 & 884  \\
      US Commute     & Transportation          & Counties         & People             & 4.31     & 2.02   & 4.3                 & 3109 \\
      E. Coli        & Biological              & Metabolites      & Common reactions   & 2.52     & 1.1    & 6.6                 & 1100 \\
      Human brain    & Biological              & Brain regions    & Connection density & 7.14     & 0.86   & 24.1                & 501  \\
    \end{tabular}
    \caption{\textbf{Overview of the considered real-world network data} Details for each dataset can be found in Methods.\label{tab_datasets}}
  \end{ruledtabular}
\end{table*}
Many models have been proposed to generate weighted networks. Among them, growing network models~\cite{Bianconi2004a,Barrat2004b,Yook2001,Kumpula2007,Antal2005,Zheng2003,Wang2005,Li2006} and the maximum-entropy class of models~\cite{Mastrandrea2014,Garlaschelli2009a,Garlaschelli2009,Sagarra2013,Sagarra2015}. However, none of them is general enough to reproduce simultaneously the topology and weighted structure of real weighted complex networks. We introduce a new model based on a class of random networks with hidden variables embedded in a metric space~\cite{Serrano08_PhysRevLett,Boguna2009} that overcomes these limitations. In this model, $N$ nodes are uniformly distributed with constant density $\delta$ in a $D$-dimensional homogeneous and isotropic metric space (see Supplementary~Methods), and are assigned a hidden variable $\kappa$ according to the probability density function (pdf) $\rho(\kappa)$. Two nodes with hidden variables $\kappa$ and $\kappa^\prime$ separated by a metric distance $d$ are connected with a probability
\begin{align} \label{eq:chi}
  \mbox{Prob}(\kappa,\kappa^\prime,d) = p(\chi), \; \; \mbox{ and } \; \; \chi = \frac{ d }{  (\mu \kappa \kappa^\prime)^{1/D} } \ ,
\end{align}
where $\mu>0$ is a free parameter fixing the average degree and $p(\chi)$ is an arbitrary positive function taking values within the interval $(0,1)$. The free parameter $\mu$ can be chosen such that $\bar{k}(\kappa) = \kappa$. Hence, $\kappa$ corresponds to the expected degree of nodes, so the degree distribution can be specified through the pdf $\rho(\kappa)$ regardless of the specific form of $p(\chi)$ (see Supplementary~Methods). The freedom in the choice of $p(\chi)$ allows us to tune the level of coupling between the topology of the networks and the metric space, which in turn allows us to control many properties such as the clustering coefficient and the navigability \cite{Serrano08_PhysRevLett,Boguna2009c}.

To generate weighted networks, a second hidden variable $\sigma$ is associated to each node. This new hidden variable can be correlated with $\kappa$ so, hereafter, we assume that the pair of hidden variables $(\kappa,\sigma)$ associated with the same node are drawn from the joint pdf $\rho(\kappa,\sigma)$. The weight of an existing link between two nodes with hidden variables $\kappa_i$, $\sigma_i$, $\kappa_j$ and $\sigma_j$, respectively, and at a metric distance $d_{ij}$ is given by
\begin{equation} \label{eq:omega}
  \omega_{ij}=\epsilon_{ij} \frac{\nu \sigma_i \sigma_j}{(\kappa_i \kappa_j)^{1-\alpha/D} d_{ij}^\alpha}
\end{equation}
with $\nu > 0$ and $0 \leq \alpha < D$ and where $\epsilon$ is a positive random variable drawn from the probability density function $f(\epsilon)$. Notice that $\alpha$ dictates a trade-off between the contribution of degrees and geometry to weights. If $\alpha=0$ weights are independent of the underlying metric space and maximally dependent on degrees, while $\alpha=D$ implies that weights are maximally coupled to the underlying metric space with no direct contribution of the degrees. Equation~\eqref{eq:omega} constitutes the keystone of our model. Indeed, as shown in the Supplementary~Methods, the form of Eq.~\eqref{eq:omega} is the only one ensuring that $\bar{s}(\sigma) \propto \sigma$. The free parameter $\nu$ can then always be chosen such that $\bar{s}(\sigma) = \sigma$.  The new hidden variable $\sigma$ can therefore be interpreted as the expected strength of a node, and the joint pdf $\rho(\kappa,\sigma)=\rho(\kappa) \rho(\sigma|\kappa)$ controls the correlation between degrees and  strengths in the network. Indeed, as shown in the Supplementary~Methods, the average strength of nodes with a given degree, $\bar{s}(k)$, relates to the first moment of the conditional pdf $\rho(\sigma|\kappa)$, $\bar{\sigma}(\kappa)$, so that when $\lim_{\kappa \rightarrow \infty}\bar{\sigma}(\kappa)=\infty$ then $\bar{s}(k) \sim \bar{\sigma}(\kappa)$. 

The relations $\bar{k}(\kappa)=\kappa$ and $\bar{s}(\sigma)=\sigma$---and consequently the relation between $\rho(\kappa, \sigma)$ and the degree-strength distribution---hold independently of the specific form of the connection probability $p(\chi)$ and of the noise distribution $f(\epsilon)$. Besides conferring great versatility to our model, this conveys a degree of control over the weight distribution which is independent of the specification of degrees and strengths and, more importantly, opens the possibility to measure the metric properties of complex weighted networks.

To use the model in the context of real weighted networks, we choose the circle $\mathbb{S}^1$ of radius $R=N/2\pi$ to be the underlying geometry, i.e. $D=1$, over which $N$ nodes are uniformly distributed~\cite{Serrano08_PhysRevLett}. Distances among nodes are measured in terms of arc lengths, that is, two nodes with angular positions $\theta$ and $\theta^\prime$ are at a distance $d(\theta,\theta^\prime)=R\Delta\theta$ where $\Delta\theta=\pi-|\pi-|\theta-\theta^\prime||$. The connection probability is set to 
\begin{align} \label{eq:illustration_prob_connection}
  p(\chi) = \frac{1}{1+\chi^\beta} \  \; \; \mbox{with}  \; \; \chi = \frac{ d }{  \mu \kappa \kappa^\prime },
\end{align}
where $\beta>1$ is a free parameter that can be used to tune the clustering and quantifies the level of coupling between the network topology and the metric space. Equation~\eqref{eq:illustration_prob_connection} casts the ensemble of networks generated by the model into exponential random networks \cite{Krioukov2010}, i.e., networks that are maximally random given the constraints imposed by the free parameters (i.e., $\rho(\kappa)$ and $\beta$). To obtain a scale-free degree distribution, hidden variables $\kappa$ are distributed according to $\rho(\kappa) \propto \kappa^{-\gamma}$ with $\kappa_0<\kappa<\kappa_\mathrm{c}$ and $\gamma>1$. 

Weights are assigned on top of the topology generated by the model. The noise distribution $f(\epsilon)$ is chosen to be a gamma distribution of average $\langle \epsilon \rangle=1$ with a given second moment $\langle \epsilon^2 \rangle$. Finally, to control the correlation between strength and degree and, therefore, to tune the strength distribution, we assume a deterministic relation between hidden variables $\sigma$ and $\kappa$ of the form $\sigma = a\kappa^\eta$, as observed in real complex networks~\cite{Barrat2004a,Popovic2012,Barthelemy11_PhysRep}, yielding $\bar{s}(k)\sim ak^\eta$ (see Supplementary~Methods). Notice that the relation between average strength and degree in the previous expression is totally independent of the underlying metric space, which implies that the strength distribution scales as $P(s)\sim s^{-\xi}$ for $s \gg 1$ with $\xi = (\gamma+\eta-1)/\eta$. All these theoretical predictions and the ones derived in the Supplementary~Methods are confirmed in Supplementary~Figure~1.
%
%
%
%
%
\subsection*{Hidden metric spaces underlying real weighted networks}
%
At the beginning of this section, we showed that the normalised weights of links participating in triangles are higher, thus suggesting a coupling between the weighted organisation of real weighted complex networks and an underlying metric space. We then presented a model that has the critical ability to fix the joint degree-strength distribution while, independently, varying the level of coupling between the weights and the metric space (parameter $\alpha$). This opens the way to a definite proof of the geometric nature of weights in real complex networks, which inevitably must involve the triangle inequality: the most fundamental property of any metric space.

For unweighted networks, a direct verification of the triangle inequality based on the topology without an embedding in a metric space is not possible, due to the probabilistic nature of the relationship between the binary structure and the distance between nodes. In contrast, weights do contain information about their distances in the metric space [via Eq.~\eqref{eq:omega}] such that a direct verification of the triangle inequality is possible. To ensure that the metric properties of triples in the network are in correspondence to the metric properties of the corresponding triangles in the underlying space, only triples of nodes forming triangles in the network are taken into account to evaluate the triangle inequality. There are however two main challenges when one tries to apply this methodology. The first one is related to the fact that connections in the weighted $\mathbb{S}^1$ model depend not only on angular distances but also on hidden degrees such that we need a purely geometrical formulation of the weighted hidden metric space network model in which angular distances and degrees are combined into a single distance measure. The second issue is related to the intrinsic noise present in the system due to the stochastic nature of the processes conforming it, which may blur the evaluation of the triangle inequality. Below, we propose a way to overcome these two issues. 

First, as shown in~\cite{Krioukov2010}, the model described by Eq.~\eqref{eq:chi} is equivalent, in the one dimensional case, to a purely geometric model where nodes are embedded within a disk of radius $R$ in the hyperbolic plane of constant curvature $-1$. Indeed, by mapping the hidden variable $\kappa$ to a radial coordinate $r$ as follows
\begin{equation} \label{mapping}
  r=R-2 \ln{\left[ \frac{\kappa}{\kappa_0}\right]}
  \; \;\mbox{ with }\; \;
  R=2\ln{\left[ \frac{N}{\mu \pi \kappa_0^2}\right]}
\end{equation}
and keeping the same angular coordinates, the connection probability Eq.~\eqref{eq:chi} can be written as
\begin{equation} \label{eq:geometric_equivalence}
  p\left( \frac{d}{\mu \kappa \kappa'} \right)=p\left( e^{\frac{1}{2} (x-R)}\right)
\end{equation}
where $x=r+r'+2 \ln{\frac{\Delta \theta}{2}}$ is a very good approximation of the hyperbolic distance between two points with radial coordinates $r$ and $r'$ and angular separation $\Delta \theta$. In this framework, networks generated with our model are geometric random networks in the hyperbolic plane, a geometry in which the triangle inequality must hold. To test the triangle inequality, we therefore select nodes participating in topological triangles in the network and measure the hyperbolic distance between them.
%
%
%
\begin{figure}[t]
  \includegraphics[width = \linewidth]{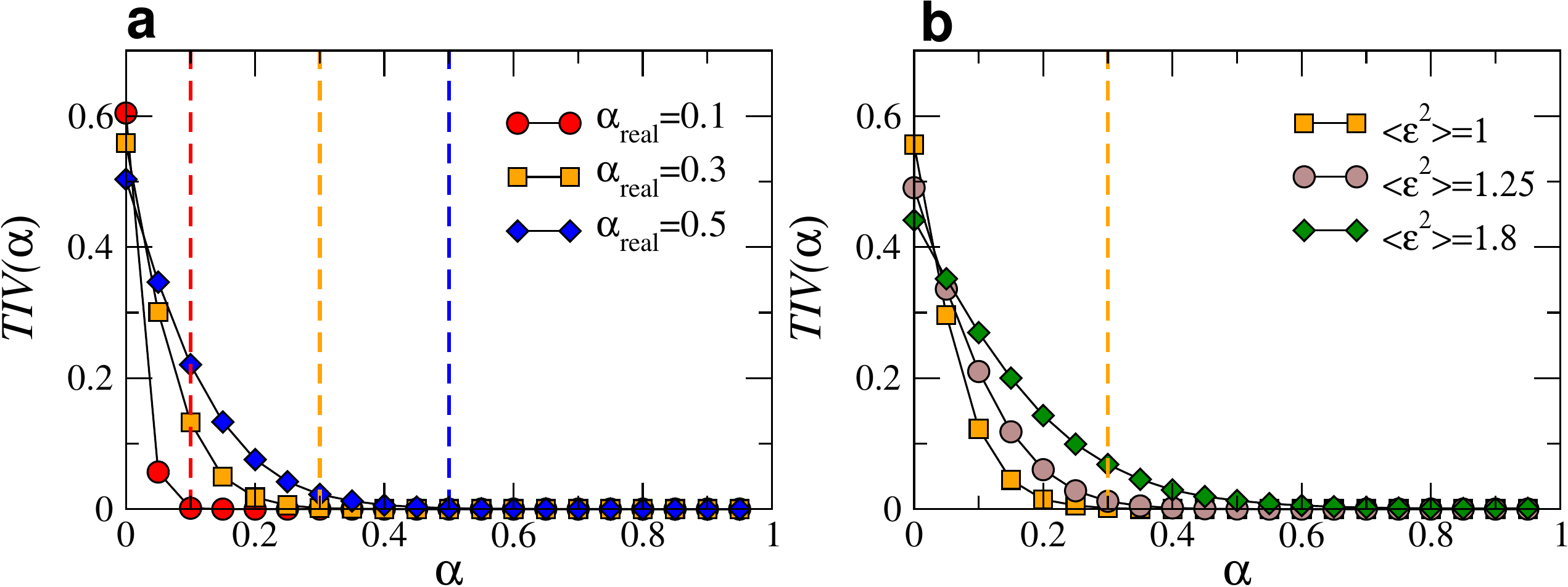}
  \caption{\textbf{Triangle inequality violation in synthetic networks. Fraction of violations of the triangle inequality}, $TIV(\alpha)$, (i.e. Eq.~\eqref{TIV} {\bf a}, without noise and different values of $\alpha_\mathrm{real}$ and {\bf b}, with a fixed value of $\alpha_\mathrm{real}=0.3$ and different values of the noise $\langle \epsilon^2 \rangle$. In both cases, the topology is the same, with $\gamma=2.5$, $\beta=2$, $\eta=1$, and $N=10^4$. The vertical dashed lines indicate the values of $\alpha_\mathrm{real}$ used to generate the different networks.}
  \label{fig:tiv}
\end{figure}

The purely geometric interpretation of our model given by Eq.~\eqref{eq:geometric_equivalence} further illustrates the reasons for which a metric space implies a non-vanishing clustering even in the thermodynamical limit. As stated at the beginning of this section, the triangle inequality---a fundamental property of any metric space, including the hyperbolic plane---stipulates that whenever point A is close to point B and point B is close to point C, then points A and C are also close. Consequently, the notion of ``closeness'' extends well beyond pairwise comparisons and is integrated ``at once'' in the positions in the metric space. This implies that many-body interactions emerge from pairwise interactions, such as the connection probability given by Eq.~\eqref{eq:illustration_prob_connection}. Given that nearby nodes are likely to be connected, clustering is a direct consequence of such many-body interactions; any triad of close nodes are likely to form a triangle, independently of the size of the disk, and therefore of the total number of nodes.

By using the mapping given by Eq.~\eqref{mapping}, with $D=1$ Eq.~\eqref{eq:omega} becomes
\begin{equation} \label{eq:weights_in_H2}
  \omega_{ij}=\epsilon_{ij} \frac{\nu}{\mu^{\alpha}} \frac{\sigma_i \sigma_j}{\kappa_i \kappa_j} e^{-\frac{\alpha}{2}(x_{ij}-R)},
\end{equation}
from which we can isolate the hyperbolic distance, $x_{ij}$, between nodes $i$ and $j$. The triangle inequality, $x_{ij}+x_{jk} \ge x_{ik}$, then becomes
\begin{equation}\label{TIV}
  \ln{\left[ \frac{\omega_{ij} \omega_{jk}}{\omega_{ik}} \left(\frac{\kappa_j}{\sigma_j}\right)^2\right]}+\ln{\left[\frac{\mu^{\alpha}}{\nu} \right]}-\frac{1}{2}\alpha R \le \ln{\left[ \frac{\epsilon_{ij} \epsilon_{jk}}{\epsilon_{ik}} \right]}.
\end{equation}
The first term in the left hand side of this inequality is a function of the actual weights and network topology and, thus, can be empirically estimated in any network. The next two terms on the left hand side have an explicit dependence on the parameter $\alpha$ (note that $\nu$ also depends on $\alpha$). The term in the right hand side is a noise term whose mean value is close to zero. 

Let us first assume that this noise term is zero. In synthetic weighted networks, the inequality should hold approximately for any value of $\alpha$ in Eq.~\eqref{TIV} equal to or larger than the value of $\alpha_\mathrm{real}$ used to assign weights in the network. Note that it may not hold exactly even when $\alpha$ is greater than its real value due to the inherent noise in the estimation of the hidden variables $\kappa$ and $\sigma$ in Eq.~\eqref{TIV}, as well as the global parameters $\mu$ and $\nu$ (note that whenever we set $\bar{s}(\sigma) = \sigma$, parameter $\nu$ becomes a function of $\alpha$, see Supplementary~Methods). To minimise such uncertainty, we choose $\sigma=a \kappa^{\eta}$ and approximate $\kappa$ by the degree of nodes. We propose to consider $\alpha$ in Eq.~\eqref{TIV} as a free parameter and to measure the triangle inequality violation spectrum, $TIV(\alpha)$, defined as the fraction of violations of the triangle inequality (i.e., triangles for which the left hand side of Eq.~\eqref{TIV} is positive). In the absence of noise, $TIV(\alpha)$ should take a very small value when $\alpha \ge \alpha_\mathrm{real}$ if the weighted structure of the network is congruent with the existence of an underlying metric space. In Fig.~\ref{fig:tiv}{\bf a}, we show $TIV(\alpha)$ for synthetic networks generated with the model with different values of $\alpha_\mathrm{real}$. As expected, the curves fall rapidly precisely at $\alpha\gtrsim\alpha_\mathrm{real}$, indicated by the dashed vertical lines.

In real situations, however, noise is typically present and has an impact on $TIV(\alpha)$. Indeed, Fig.~\ref{fig:tiv}{\bf b} shows its behaviour for a fixed value of $\alpha_\mathrm{real}$ and different values of the noise $\langle \epsilon^2 \rangle$. This implies that we need an independent measure of the noise to infer the value of $\alpha_\mathrm{real}$ from the spectrum $TIV(\alpha)$. For this purpose, we use the square of the coefficient of variation of the strength, which depends linearly on the noise $\langle \epsilon^2 \rangle$ (see the Supplementary~Methods). Combining these observations, we propose a procedure to infer the value of $\alpha_\mathrm{real}$ for any real complex network based on the empirical $TIV(\alpha)$. The method is described in details in the Supplementary~Methods.
  
Figures~\ref{fig:real_alpha}{\bf a} and {\bf b} show the $TIV(\alpha)$ curves for the real networks and the same curves for synthetic networks generated by our model using the inferred $\alpha_\mathrm{real}$ to be maximally congruent with the real data. In all cases, we find a very good agreement between theory and observations, which suggests a coupling with a hidden metric space as a highly plausible explanation of the observed weighted organisation. Note that the increase of $TIV(\alpha)$ for $\alpha \sim 1$ is an expected artifact of Eq.~\eqref{TIV} (see Supplementary~Methods). Figure~\ref{fig:real_alpha}{\bf c} shows the values of $\beta$ (coupling topology and metric space) and $\alpha_\mathrm{real}$ (coupling weights and metric space) inferred by our method. Notice that, except for the US airports network, $\alpha_\mathrm{real}$ is always above 0.40, which indicates a clear and strong coupling between weights and the hidden underlying geometry. We also generated synthetic networks with the inferred parameters and confronted their topological and weighted properties against those of their real counterparts (see Fig.~\ref{fig:iJ01366} and Supplementary~Methods for other networks and a comparison with other models). In all cases, the agreement between the model and the real networks is excellent. Remarkably, in the case of the weight distribution and disparity measure, such agreement is only achieved with the empirical value of $\alpha_\mathrm{real}$ found via the test of the triangle inequality.

Finally, we considered the networks for which an embedding of the binary structure was available and rescaled each weight by a factor $e^{-\alpha_\mathrm{real} x_{ij} / 2}$ [see Eq.~\eqref{eq:weights_in_H2}] where $x_{ij}$ is the hyperbolic distance between nodes in the embeddings \cite{Boguna10_NatCommun}. We then normalised and sampled the weights as in Fig.~\ref{fig:1} and the results are shown in Fig.~\ref{fig:real_alpha}{\bf d}. Strikingly, we see that the gap observed in Fig.~\ref{fig:1} completely disappears in some of the networks or is significantly reduced in others. While the remaining gaps may be due to imprecisions in the embedding (i.e. the embedding procedure cannot take into account the information contained in the weights yet), these results nevertheless add their voice to the evidence pointing toward the geometric nature of the weights in real complex networks.
%
%
%
\begin{figure}[t]
\includegraphics[width = \linewidth]{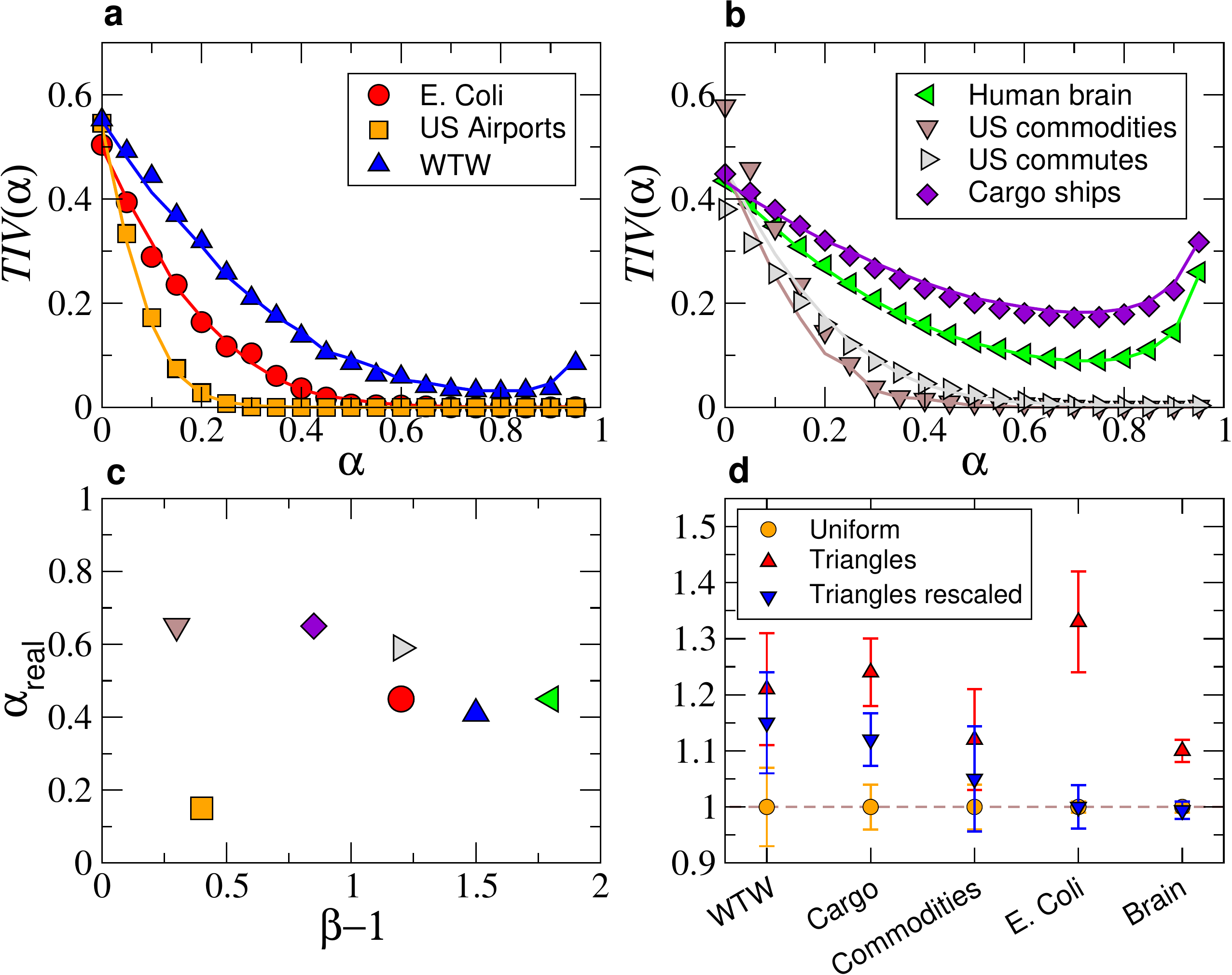}
\caption{\textbf{Triangle inequality violation in real networks.} Triangle inequality violation curves ({\bf a} and {\bf b}) for all real networks considered in this study (symbols). Solid lines correspond to their model counterparts with the model parameters in Supplementary~Table~1. {\bf c}, Inferred values of the coupling parameter $\alpha_\mathrm{real}$ vs. $\beta-1$. The numerical values of these parameters can be found in Supplementary~Table~1. \textbf{d}, Average normalised weights in the network (yellow circles) with the one measured by sampling links over triangles (red triangles) for most of the empirical datasets analysed. The error bars correspond to an estimate of the standard deviation of the average value due to the finite size of the samples (see the caption of Fig.~\ref{fig:1} for details). The blue inverted triangles correspond to the red triangles but where the weights were rescaled by the factor $e^{-\alpha_\mathrm{real} x_{ij} / 2}$ to take into account the coupling between the weights and the hidden metric space. The airports and commuting networks could not be embedded into a metric space using current state-of-the-art methodology \cite{Boguna10_NatCommun,Papadopoulos2015} due to atypical topological features and are therefore not reproduced here. These atypical features refer to a power-law degree distribution with an exponent below 2 in the case of the U.S. airports network and a short-range repulsion effect in the connection probability for the commute network (i.e., people rarely commute from one suburb to another but rather commute from one suburb to the major city in the area). This does not affect our general theory but rather prevent the state-of-the-art embedding algorithms to provide us with an embedding of these two networks.}
\label{fig:real_alpha}
\end{figure}
%
%
%
\begin{figure}[t]
\includegraphics[width = \linewidth]{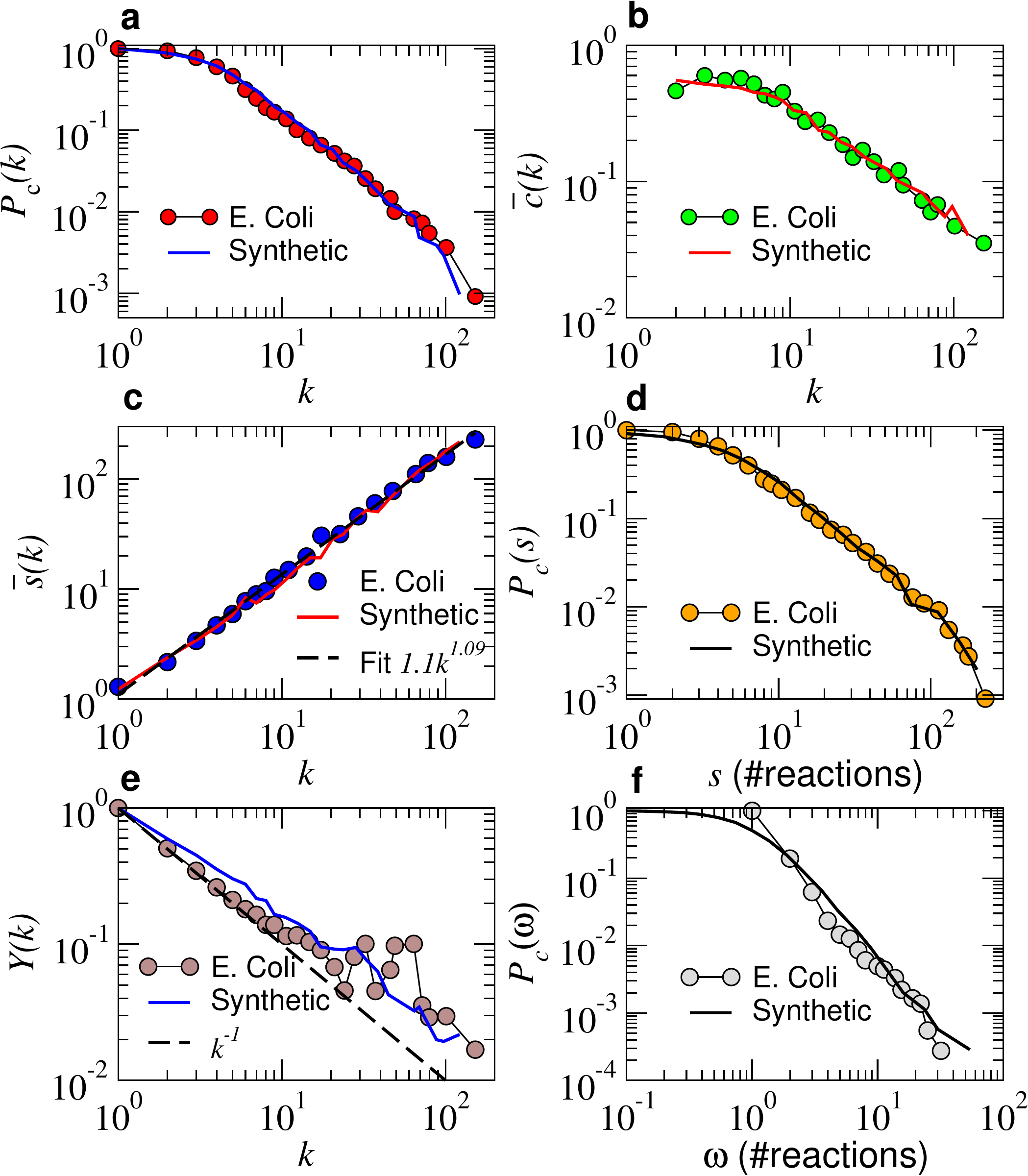}
\caption{\textbf{Model versus real network.} Comparison between topological and weighted properties of the iJO1366 {\it E. Coli} metabolic network (symbols) and a synthetic network generated by the model with the parameters given in Supplementary~Table~1 (solid lines). {\bf a}, Complementary cumulative degree distribution. {\bf b}, Degree-dependent clustering coefficient.  {\bf c}, Average strength of nodes of degree $k$.  {\bf d}, Complementary cumulative strength distribution. {\bf e}, Disparity of nodes as a function of their degree (see Methods). {\bf f}, Complementary cumulative weight distribution of links.}
\label{fig:iJ01366}
\end{figure}
%
%
%
%
%
\section*{Discussion}
%
The metric character of many real complex networks---in which clustering is a direct consequence of the triangle inequality---has long been established. However, the metric nature of their weighted organisation still remained an open question. In this paper, we provided strong empirical evidence for the metric origin of the weighted architecture of real complex networks from very different domains. Our results suggest that the same underlying metric space ruling the network topology also shapes its weighted organisation. It is important to notice that the distances between nodes implied by this metric space does not necessarily correspond to geographic distances (e.g., distances between ports on the Earth) but are rather abstract and effective distances encoding several factors affecting the existence of connections and their intensity.

To account for these empirical findings, we proposed a very general model capable of reproducing the coupling with the metric space in a very simple and elegant way. This model allows us to fix the local properties of the nodes---their joint degree-strength distribution---while varying the coupling of the topology and, independently, of the weights with the hidden metric space. This critical property permits us to gauge quantitatively the effect of the metric space in real systems. In the case of the US airports network, we found quite remarkably that while the coupling between the topology and the metric space is relatively strong, the coupling at the weighted level is quite weak. This strengthens the hypothesis that in some systems the formation of weights and topology obey different dynamics. Contrarily, we found strong coupling, both at the topological and weighted levels, even in networks that are not embedded in any obvious metric space like the metabolism of  \textit{E. Coli}, a system of metabolic reactions for which the hidden geometry is elucidated as a biochemical affinity space. This fact provides yet another empirical evidence towards the existence of hidden metric spaces shaping the architecture of these systems and, more generally, of real complex networks~\cite{Serrano08_PhysRevLett}.

Our framework can be understood as a new generation of gravity models applicable to very different domains, including Biology, Information and Communication Technologies, and Social Systems. Indeed, Eq.~\eqref{eq:omega} is a novel generalisation of this concept to the case of weighted networks where
\begin{align}
  \frac{\sigma}{\kappa^{1-\alpha/D}}
\end{align}
plays the role of the ``mass'' of nodes and ensures that, once the network has been assembled, nodes have expected degree and strength $\kappa$ and $\sigma$, respectively. Current gravity models predict the volume of flows between elements but cannot explain the observed topology of the interactions among them, as shown in works for the world trade web~\cite{Duenas2013}. Our contribution overcomes this limitation and offers a gravity model that can reproduce both the existence and the intensity of interactions. This opens a new line of theoretic research on the coupling between topology, weighted structure, and geometry in complex networks.

Furthermore, our work opens the possibility to use information encoded in the weights of the links to find more accurate embeddings of real networks. Such improved embeddings are expected to allow the detection of communities or of missing links and to provide estimates of the weights of such missing links~\cite{Liben-Nowell:2007aa,Lu:2010aa,Zhao:2015aa}. They can also be extremely helpful to implement navigation and searching protocols, such as greedy routing, which take into account not only the existence of connections but also their intensity.

In perspective, the hidden metric space weighted model and the maps of real complex systems that it will enable will lead to a deeper understanding of the interplay between the structure and function of real networks, and will offer insights on the impact they have on the dynamical processes they support and on their own evolutionary dynamics.
%
%
%
%
%
\appendix
\section*{Methods}
%
%
%
\subsection*{Empirical datasets}
%
In addition to the details given in Table~\ref{tab_datasets}, we provide further information and references about the real complex networks used in this paper.

The world trade web describes significant trade exchanges between countries in 2013. The corresponding weights are trade volumes between pairs in USD~\cite{Garcia-Perez2016}.

The international network of global cargo ship movements consists of the number of shipping journeys between pairs of major commercial ports in the world in 2007~\cite{Kaluza2010}.

The commodities network corresponds to the flows of the goods and services in millions of USD between industrial sectors in the United States in 2007\cite{Grady2012}.

The airports network indicates the number of passengers that flew between pairs of airports in the United States in 2013. Data is freely available at the website of the U.S. Bureau of Transportation Statistics (\texttt{transtats.bts.gov}).

The commuting network reflects the daily flow of commuters between counties in the United States in 2000~\cite{Grady2012}.

Weights in the metabolic network of the bacteria \textit{Escherichia Coli (E. Coli)} K-12 MG1655 consist of the number of different metabolic reactions in which two metabolites participate~\cite{Orth:2011a,Serrano2012}.

Weights in the human brain network correspond to the density of anatomical connections between subregions of the human brain as detected via diffusion tensor imaging~\cite{Nicholas2010}.

Except for the metabolic and human brain networks, all networks were filtered using the disparity filter defined in Ref.~\cite{Serrano2009} to preserve the most statistically significant connections. Many real weighted networks are generated from data by using a very broad definition of what constitutes a significant connection. This results in networks with huge average degrees and in which many links are noisy and weakly related to the overall functionality of the network. For instance, the US airports network contains links due to private flights (of the order of 10 passengers per year), which obviously follow different patterns of connection than the regular commercial airlines. Another interesting example is the world trade web, in which many trade interactions amount for less than one million dollars and are extremely volatile, appearing and disappearing from year to year. Indeed, it has been shown in Ref.~\cite{Garcia-Perez2016} that removing these noisy connections yields a significantly more congruent topology with real economic factors, such as the gross domestic product (GDP).
%
%
%
\subsection*{Disparity}
%
The disparity quantifies the local heterogeneity of the weights attached to a given node and is defined as
\begin{align}
  Y_i = \sum_{j} \left( \frac{\omega_{ij}}{s_i} \right)^2 \ ,
\end{align}
where $\omega_{ij}$ is the weight of the link between nodes $i$ and $j$ ($\omega_{ij}=0$ if there is no link) and $s_i=\sum_j \omega_{ij}$ \cite{Barthelemy2005}. From this definition, we see that the disparity scales as $Y_i \sim k_i^{-1}$ whenever the weights are roughly homogeneously distributed among the links. Conversely, whenever the disparity decreases slower than $k_i^{-1}$ implies that weights are heterogeneous and that the large strength of a node is due to a handful of links with large weights.
%
%
%
%
%
%
%

%
%
%
%
%
%
\begin{acknowledgments}
  We acknowledge support from the James S. McDonnell Foundation Scholar Award in Complex Systems; the Fonds de recherche du Qu\'ebec -- Nature et technologies; the ICREA Academia prize, funded by the Generalitat de Catalunya; the MINECO project no.FIS2013-47282-C2-1-P; and the Generalitat de Catalunya grant no.2014SGR608.
\end{acknowledgments}
\noindent\textbf{Author contributions}: A.A., M.\'A.S., G.G.-P., and M.B. contributed to the design and implementation of the research, to the analysis of the results, and to the writing of the manuscript.\vspace{\baselineskip}\\
\textbf{Competing financial interests}: The authors declare no competing financial interests.\vspace{\baselineskip}\\
\textbf{Data availability}: Codes and data supporting the findings of this study are available from the corresponding author upon request.
\clearpage


\section*{Supplementary material}

# Supplementary Methods

Further information and details about the model and the methods described in the main test are provided. Section I details the analytical results that can be obtained for the model presented in the paper. For the sake of completeness, some details discussed in the main text are reproduced. Section II discusses the method used to infer the value of $\alpha$ via the triangle inequality test. Section III compares the predictions of our model with other models proposed in the literature.

## I. FURTHER DETAILS ON THE MODEL

Our model is a non-trivial generalization to weighted networks of a class of random networks with hidden variables embedded in a metric space [1]. In this model, $N$ nodes are uniformly distributed with density $\delta$ in a $D$-dimensional homogeneous and isotropic metric space, and are assigned a hidden variable $\kappa$ according to the probability density function (pdf) $\rho(\kappa)$. Two nodes with hidden variables $\kappa$ and $\kappa'$ separated by a metric distance $d$ are connected with a probability

$$p(\kappa, \kappa', d) = p(\chi) \quad \text{where} \quad \chi = \frac{d}{(\mu\kappa\kappa')^{1/D}} \ , \tag{1}$$

where $\mu > 0$ is a free parameter and $p(\chi)$ is an arbitrary positive function taking values within the interval $(0, 1)$. Whenever the integral $\int_0^\infty \chi^{D-1} p(\chi) d\chi$ is bounded, the free parameter $\mu$ can be chosen such that $\bar{k}(\kappa) = \kappa$. Hence, $\kappa$ corresponds to the expected degree of nodes, so the degree distribution can be specified through the pdf $\rho(\kappa)$

$$P(k) = \frac{1}{k!} \int e^{-\kappa} \kappa^k \rho(\kappa) d\kappa \ , \tag{2}$$

regardless of the specific form of $p(\chi)$ (see Secs. I A 1 and I A 2 below for details). The freedom in the choice of $p(\kappa, \kappa', d)$ allows us to tune the level of coupling between the topology of the networks and the metric space, which in turn allows us to control many properties such as the clustering coefficient and the navigability [1, 2]. Moreover, the form of the connection probability in Eq. (1) implies that networks generated with this model are small worlds for any heterogeneous pdf $\rho(\kappa)$ since high degree nodes are then likely to be connected regardless of the metric distance between them [1, 3].

To generalize this model to weighted networks, a second hidden variable $\sigma$ is associated to each node. This new hidden variable can be correlated with $\kappa$ so, hereafter, we assume that the pair of hidden variables $(\kappa, \sigma)$ associated with the same node are drawn from the joint pdf $\rho(\kappa, \sigma)$. The weight of an *existing* link between two nodes with hidden variables $\kappa$, $\sigma$, $\kappa'$ and $\sigma'$, respectively, and at a metric distance $d$ is distributed according to the pdf

$$\phi(w|\kappa, \sigma, \kappa', \sigma', d) = \frac{1}{\bar{w}} f\left(\frac{w}{\bar{w}}\right) \ , \tag{3}$$

where $f(\epsilon)$ is any probability density function in the domain $[0, \infty)$, and where

$$\bar{w} = \frac{\nu\sigma\sigma'}{(\kappa\kappa')^{1-\alpha/D} d^\alpha} \ , \tag{4}$$

with $\nu > 0$ and $0 \le \alpha < D$. The particular form of the distribution of weights Eq. (3) implies that the weight between nodes $i$ and $j$ can be written as

$$\omega_{ij} = \epsilon_{ij} \frac{\nu\sigma_i\sigma_j}{(\kappa_i\kappa_j)^{1-\alpha/D} d_{ij}^\alpha} \tag{5}$$

where $\epsilon_{ij}$ is a random variable drawn from the pdf function $f(\cdot)$. Equations (3) and (4) constitute the keystone of our model. Indeed, as shown in Sec. I A 3 below, the form of Eq. (4) is the only ensuring that $\bar{s}(\sigma) = \sigma$, provided that the integral $\int_0^\infty \chi^{D-\alpha-1} p(\chi) d\chi$ converges. The new hidden variable $\sigma$ can therefore be interpreted as the expected strength of a node, and the joint pdf $\rho(\kappa, \sigma) = \rho(\kappa)\rho(\sigma|\kappa)$ controls the correlation between degrees and strengths in the network. Indeed, as shown in Sec. I A 4 below, the average strength of nodes with a given degree, $\bar{s}(k)$, relates to the first moment of the conditional pdf $\rho(\sigma|\kappa)$, $\bar{\sigma}(\kappa)$, through the relation

$$\bar{s}(k) = \frac{1}{(k-1)!P(k)} \int e^{-\kappa} \kappa^{k-1} \rho(\kappa)\bar{\sigma}(\kappa)d\kappa \ . \tag{6}$$

Therefore, when $\lim_{\kappa\to\infty} \bar{\sigma}(\kappa) = \infty$ then $\bar{s}(k) \sim \bar{\sigma}(\kappa)$. This limit stands as a good approximation for the behaviour of high degree nodes in real weighted networks.

Remarkably, these relations hold independently of the specific form of the connection probability $p(\chi)$ and of the distribution of weights $f(\epsilon)$, thus conferring great versatility to our model. Even more remarkable, the shape of the connection probability $p(\chi)$ and the value of the parameter $\alpha$—coupling topology and weights to the metric space—do not affect the relations $\bar{k}(\kappa) = \kappa$ and $\bar{s}(\sigma) = \sigma$ and, therefore, the join degree-strength distribution $P(k, s)$. This property conveys a degree of control over the weight distribution as well as over the disparity of nodes which is independent of the specification of degrees and strengths and, more importantly, opens the possibility to measure the metric properties of complex weighted networks.

### A. Theoretical calculations in a $D$-dimensional metric space

Most of the theoretical calculations for the model are carried out using the probability density function (pdf) $g(k, s|\kappa, \sigma)$ corresponding to the probability that a node with hidden variables $\kappa$ and $\sigma$ has a degree equal to $k$ and a strength in the interval $[s, s + ds]$. To compute this pdf, we first consider a pair of nodes whose positions in space are $\boldsymbol{x_i}$ and $\boldsymbol{x_j}$ and whose hidden variables are $\kappa_i$, $\sigma_i$ and $\kappa_j$, $\sigma_j$, respectively. The pdf for the weight $w_{ij}$ between these two nodes ($w_{ij} = 0$ means that there is no link) is

$$\Phi(w_{ij}|\kappa_i, \sigma_i; \kappa_j, \sigma_j; d_{ij}) = [1 - p(\kappa_i, \kappa_j, d_{ij})]\delta(w_{ij}) \\ + p(\kappa_i, \kappa_j, d_{ij})\phi(w_{ij}|\kappa_i, \sigma_i, \kappa_j, \sigma_j, d_{ij})\Theta(w_{ij}) \ , \tag{7}$$

where $\delta(\cdot)$ is the Dirac delta function and $\Theta(\cdot)$ is the left-continuous Heaviside step function [i.e., $\Theta(0) = 0$]. Without loss of generality, we take the perspective of node $i$, place it at the origin of the coordinate system, and integrate over the possible values of the hidden variables of node $j$

$$\Phi(w_{ij}|\kappa_i, \sigma_i) = \iiint \frac{\rho(\kappa_j, \sigma_j)}{V_D} \Phi(w_{ij}|\kappa_i, \sigma_i; \kappa_j, \sigma_j; d_{ij})d\boldsymbol{x_j}d\sigma_j d\kappa_j \ . \tag{8}$$

In this last equation, $1/V_D$ is the (uniform) pdf for the position of nodes in the metric space. This last expression does not depend on the position of node $i$ due to the isotropy, homogeneity and large size ($N \gg 1$) of the metric space. Note that for mathematical convenience, we consider the metric space to be a $D$-dimensional sphere of radius $R$. However, the constraint of having constant density implies that the radius diverges in the thermodynamic limit and, thus, the metric space is equivalent to a $D$-dimensional Euclidean space. Because the weights $\{w_{ij}\}_{i,j=1,...,N}$ are assigned independently in the model, the probability that a node has a degree $k_i$ and a strength $s_i$ is equal to the product of the contribution of each potential $N - 1$ links, given that the number of existing

links is equal to $k_i$ and that their weights sum up to $s_i$

$$g(k_i, s_i | \kappa_i, \sigma_i) = \prod_j \left[ \int \Phi(w_{ij} | \kappa_i, \sigma_i) dw_{ij} \right] \delta \left( k_i - \sum_l \Theta(w_{il}) \right) \delta \left( s_i - \sum_l w_{il} \right) . \tag{9}$$

Although it is not possible to further the calculation and obtain a closed form the the pdf $g(k, s | \kappa, \sigma)$, many useful results can be obtained using its generating function defined as

$$\hat{g}(x, y | \kappa_i, \sigma_i) = \sum_{k_i=0}^{N-1} \int_0^\infty g(k_i, s_i | \kappa_i, \sigma_i) x^{k_i} e^{-s_i y} ds_i , \tag{10}$$

which, dropping the subscripts $i$ and $j$, can be written as

$$\hat{g}(x, y | \kappa, \sigma) = \Bigg[ \iiint \frac{\rho(\kappa', \sigma')}{V_D} \Bigg\{ [1 - p(\kappa, \kappa', d)]$$
$$+ p(\kappa, \kappa', d) x \int \phi(w | \kappa, \sigma, \kappa', \sigma', d) e^{-wy} dw \Bigg\} d\boldsymbol{x'} d\sigma' d\kappa' \Bigg]^{N-1} . \tag{11}$$

### 1. The degree of nodes

From Eq. (11), we can readily see that the average degree of nodes with hidden variables $\kappa$ and $\sigma$ only depends on $\kappa$ [4]

$$\begin{aligned} \bar{k}(\kappa, \sigma) &= \left. \frac{\partial \hat{g}(x, y | \kappa, \sigma)}{\partial x} \right|_{x=1, y=0} \\ &= \frac{N-1}{V_D} \iiint \rho(\kappa', \sigma') p(\kappa, \kappa', d) d\boldsymbol{x'} d\sigma' d\kappa' \\ &= \delta \iint \rho(\kappa') p(\kappa, \kappa', d) d\boldsymbol{x'} d\kappa' , \end{aligned} \tag{12}$$

where $\rho(\kappa)$ is the marginal pdf of $\rho(\kappa, \sigma)$, and where $\delta \equiv N/V_D$ is the density of nodes in the metric space (we consider here that $N \gg 1$). Using the definition of $p(\kappa, \kappa', d)$ given in Eq. (1) and switching to $D$-dimensional spherical coordinates ($\Omega_D$ is the solid angle subtended by a $D$-dimensional object), this last equation becomes

$$\begin{aligned} \bar{k}(\kappa) &= \delta \iiint \rho(\kappa') p(\chi) r^{D-1} dr d\Omega_D d\kappa' \\ &= \frac{2\pi^{D/2} \delta \mu}{\Gamma(D/2)} \kappa \int \kappa' \rho(\kappa') d\kappa' \int \chi^{D-1} p(\chi) d\chi \\ &= \frac{2\pi^{D/2} \delta \mu \langle \kappa \rangle I_1}{\Gamma(D/2)} \kappa , \end{aligned} \tag{13}$$

where we have noted $\langle \kappa \rangle = \int \kappa \rho(\kappa) d\kappa$ and $I_1 = \int \chi^{D-1} p(\chi) d\chi$. The average degree for the whole network is

$$\langle k \rangle = \int \bar{k}(\kappa) \rho(\kappa) d\kappa = \frac{2\pi^{D/2} \delta \mu I_1}{\Gamma(D/2)} \langle \kappa \rangle^2 . \tag{14}$$

Consequently, we see that the free parameter $\mu$ can be chosen such that $\bar{k}(\kappa) = \kappa$ and $\langle k \rangle = \langle \kappa \rangle$, that is

$$\mu = \frac{\Gamma(D/2)}{2\pi^{D/2}\delta I_1 \langle \kappa \rangle} \ . \tag{15}$$

The degree distribution of the networks generated by the model can therefore be controlled through the pdf $\rho(\kappa)$. Following similar steps, we also find that

$$\mathrm{Var}[k(\kappa)] = \bar{k}(\kappa) \ , \tag{16}$$

which implies that

$$\frac{\sqrt{\mathrm{Var}[k(\kappa)]}}{\bar{k}(\kappa)} = \frac{1}{\sqrt{\kappa}} \ , \tag{17}$$

where we used $\bar{k}(\kappa) = \kappa$. In other words, nodes with a same high value of their hidden variable $\kappa$ tend to all have a degree close to the expected value $\bar{k}(\kappa) = \kappa$.

### 2. The degree distribution

By definition, evaluating Eq. (11) at $y = 0$ yields the generating function for the degree distribution of nodes with hidden variable $\kappa$

$$\hat{g}(x,0|\kappa,\sigma) \equiv \sum_k g(k|\kappa)x^k \ . \tag{18}$$

Assuming $N \gg 1$ and using Eq. (12), we obtain

$$\begin{aligned}
\hat{g}(x,0|\kappa,\sigma) &= \left[1 + (x-1)\frac{\bar{k}(\kappa)}{N-1}\right]^{N-1} \\
&= \exp\left\{(x-1)\bar{k}(\kappa)\right\} \\
&= \sum_k \frac{\mathrm{e}^{-\bar{k}(\kappa)}\bar{k}(\kappa)^k}{k!}x^k \ .
\end{aligned} \tag{19}$$

In other words, the degrees of nodes with hidden variable $\kappa$ follow a Poisson distribution with mean $\bar{k}(\kappa)$ [4]. Using $\bar{k}(\kappa) = \kappa$, the degree distribution of the whole network is therefore

$$\begin{aligned}
P(k) &= \int g(k|\kappa)\rho(\kappa)d\kappa \\
&= \frac{1}{k!}\int \mathrm{e}^{-\kappa}\kappa^k\rho(\kappa)d\kappa \ ,
\end{aligned} \tag{20}$$

thus unveiling the precise link between $\rho(\kappa)$ and $P(k)$.

### 3. The strength of nodes

Following similar steps as in Sec. I A 1 above, we can compute the average strength of nodes with hidden variables $\kappa$ and $\sigma$

$$
\begin{aligned}
\bar{s}(\kappa,\sigma) &= -\left.\frac{\partial \hat{g}(x,y|\kappa,\sigma)}{\partial y}\right|_{x=1,y=0} \\
&= \delta \iiint \rho(\kappa',\sigma')p(\kappa,\kappa',d) \int w\phi(w|\sigma,\sigma',d)dwd\boldsymbol{x}'d\sigma'd\kappa' \\
&= \frac{2\pi^{D/2}\delta\mu^{1-\alpha/D}\nu I_2}{\Gamma(D/2)}\sigma \int \sigma'\rho(\sigma')d\sigma' \int \chi^{D-\alpha-1}p(\chi)d\chi \\
&= \frac{2\pi^{D/2}\delta\mu^{1-\alpha/D}\nu I_2\langle\sigma\rangle I_3}{\Gamma(D/2)}\sigma \ ,
\end{aligned}
\tag{21}
$$

where $\rho(\sigma)$ is the other marginal pdf of $\rho(\kappa,\sigma)$, and where we have noted $\langle\sigma\rangle = \int \sigma\rho(\sigma)d\sigma$, $I_2 = \int wf(w)dw$, and $I_3 = \int \chi^{D-\alpha-1}p(\chi)d\chi$. We therefore conclude that the strength of nodes only depends on $\sigma$, hence $\bar{s}(\kappa,\sigma) = \bar{s}(\sigma)$. The average strength for the whole network is

$$
\begin{aligned}
\langle s\rangle &= \int \bar{s}(\sigma)\rho(\sigma)d\sigma \\
&= \frac{2\pi^{D/2}\delta\mu^{1-\alpha/D}\nu I_2 I_3}{\Gamma(D/2)}\langle\sigma\rangle^2 \ .
\end{aligned}
\tag{22}
$$

As for the average degree, we see that we can set the free parameter $\nu$ such that $\bar{s}(\sigma) = \sigma$ and $\langle s\rangle = \langle\sigma\rangle$, that is

$$
\nu = \frac{\Gamma(D/2)}{2\pi^{D/2}\delta\mu^{1-\alpha/D}I_2 I_3\langle\sigma\rangle} \ .
\tag{23}
$$

The strength distribution of the networks generated by the model can therefore be controlled through the pdf $\rho(\sigma)$. Following similar steps, we find that

$$
\begin{aligned}
\mathrm{Var}[s(\kappa,\sigma)] &= \delta \iiint \rho(\kappa',\sigma')p(\kappa,\kappa',d) \int w^2\phi(w|\sigma,\sigma',d)dwd\boldsymbol{x}'d\sigma'd\kappa' \\
&= \frac{2\pi^{D/2}\delta\mu^{1-2\alpha/D}\nu^2 I_4\langle\sigma^2/\kappa\rangle I_5}{\Gamma(D/2)}\frac{\sigma^2}{\kappa} \ ,
\end{aligned}
\tag{24}
$$

where we have noted $\langle\sigma^2/\kappa\rangle = \iint(\sigma^2/\kappa)\rho(\kappa,\sigma)d\sigma d\kappa$, $I_4 = \int w^2f(w)dw$ and $I_5 = \int \chi^{D-2\alpha-1}p(\chi)d\chi$. Setting $\mu$ such that $\langle k\rangle = \langle\kappa\rangle$ [see Eq. (14)], we find that

$$
\frac{\sqrt{\mathrm{Var}[s(\kappa,\sigma)]}}{\bar{s}(\sigma)} = \frac{\sqrt{\bar{\kappa}\langle\sigma^2/\kappa\rangle I_1 I_4 I_5}}{\langle\sigma\rangle I_2 I_3}\frac{1}{\sqrt{\kappa}} \ ,
\tag{25}
$$

i.e., the strength of high-degree nodes is close to its expected value given by Eq. (21).

### 4. The strength of nodes of degree k

Unfortunately, it is not possible to obtain a general closed form of the pdf $g(s|\kappa,\sigma)$, similar to Eq. (19), from Eq. (11). We can however characterize the strengths of nodes through the average

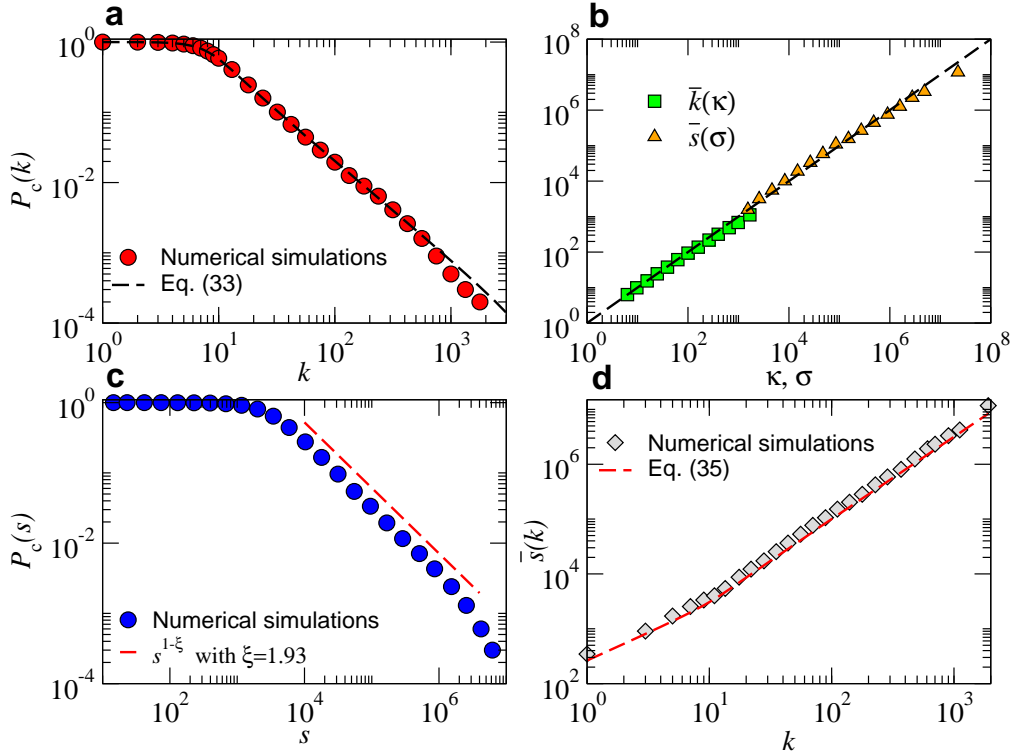

SUPPLEMENTARY FIGURE 1. **Validation of the theoretical calculations** The theoretical calculations are given in Sec. I B, and the numerical simulations correspond to a single network of size $N = 10^4$ and parameters $\alpha = 0.4$, $\beta = 1.5$, $\gamma = 2.4$, $\eta = 1.5$, $\langle k \rangle = 20$, $\kappa_c = \kappa_0 N^{1/(\gamma-1)}$, $a = 100$, and noise $\langle \epsilon^2 \rangle = 1.5$. **a**, complementary cumulative degree distribution compared with the prediction given by Eq. (33). **b**, average degree and average strength of nodes as a function of their hidden variables $\kappa$ and $\sigma$. The deviation for high $\kappa$ and $\sigma$ is due to the finite size of the network and disappears as $N \to \infty$. **c**, complementary cumulative strength distribution. The dashed line indicates a scaling $\propto s^{-\xi}$ with $\xi = (\gamma + \eta - 1)/\eta \simeq 1.93$, as expected. **d**, average strength as a function of degree. The dashed line shows the prediction of Eq. (35).

strength of nodes with a given degree, $\bar{s}(k)$. Let us first explicit its calculation

$$
\begin{aligned}
\bar{s}(k) &= \int s g(s|k) ds \\
&= \frac{1}{P(k)} \int s g(k,s) ds \\
&= \frac{1}{P(k)} \iint \left[ \int s g(k,s|\kappa,\sigma) ds \right] \rho(\kappa,\sigma) d\sigma d\kappa \ .
\end{aligned}
\tag{26}
$$

From Eq. (10), we see that the integral in brackets can be obtained from Eq. (11)

$$
-\left. \frac{\hat{g}(x,y|\kappa,\sigma)}{\partial y} \right|_{y=0} = \sum_k \left[ \int s g(k,s|\kappa,\sigma) ds \right] x^k \ .
\tag{27}
$$

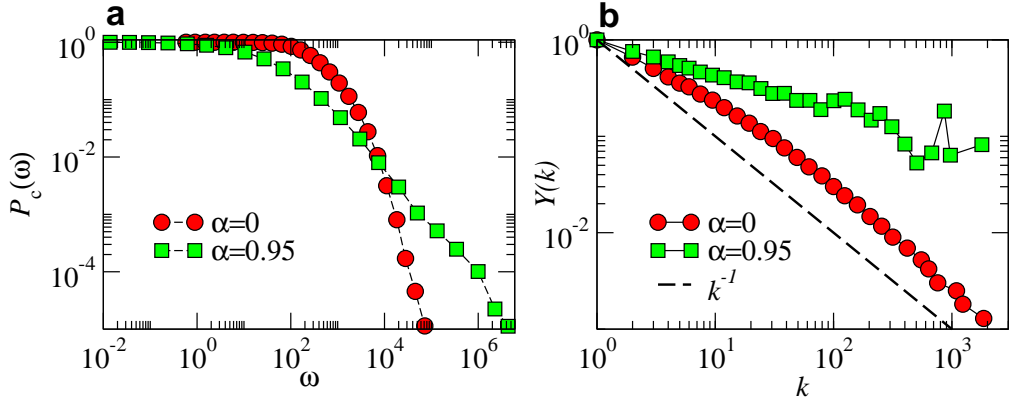

SUPPLEMENTARY FIGURE 2. **Effect of the underlying geometry on the weights** Simulations correspond to synthetic networks with the same parameters as in Supplementary Figure 1 but with two different values of the coupling parameter $\alpha$. **a**, complementary cumulative weight distribution for $\alpha = 0$ (no coupling with the hidden metric space) and $\alpha = 0.95$ (strong coupling). **b**, average disparity of nodes as a function of their degree for the same values of the coupling $\alpha$. The dashed line shows the scaling $k^{-1}$, corresponding to the a perfect equipartition of the strength of a node among its links.

Assuming $N \gg 1$ and using Eqs. (13) and (21), we find

$$
\begin{aligned}
-\left.\frac{\partial \hat{g}(x, y | \kappa, \sigma)}{\partial y}\right|_{y=0} &= x \bar{s}(\sigma)\left[1+(x-1) \frac{\bar{k}(\kappa)}{N-1}\right]^{N-2} \\
&= x \bar{s}(\sigma) \exp \left\{(x-1) \bar{k}(\kappa)\right\} \\
&= \sum_{k>0}\left[\frac{\bar{s}(\sigma) \mathrm{e}^{-\bar{k}(\kappa)} \bar{k}(\kappa)^{k-1}}{(k-1)!}\right] x^{k} .
\end{aligned}
\tag{28}
$$

Excluding the case $k = 0$ in this last expression is not problematic since $g(s|0) = \delta(s)$, by definition, and therefore the coefficient in front of $x^0$ must be zero [see Eq. (27)]. Combining Eqs. (26)–(28), $\bar{k}(\kappa) = \kappa$ and $\bar{s}(\sigma) = \sigma$, we obtain

$$
\begin{aligned}
\bar{s}(k) &= \frac{1}{(k-1)! P(k)} \iint \sigma \mathrm{e}^{-\kappa} \kappa^{k-1} \rho(\kappa, \sigma) d\sigma d\kappa \\
&= \frac{1}{(k-1)! P(k)} \int \mathrm{e}^{-\kappa} \kappa^{k-1} \rho(\kappa) \bar{\sigma}(\kappa) d\kappa .
\end{aligned}
\tag{29}
$$

thus further clarifying how the joint pdf $\rho(\kappa, \sigma)$ controls the correlation between the degree and the strength of nodes. In fact, assuming no correlation, i.e., $\rho(\kappa, \sigma) = \rho(\kappa) \rho(\sigma)$, yields $\bar{s}(k) = \langle \sigma \rangle$, which is independent of the degree, as expected.

### B. Validation of the theoretical calculations

All results presented in the previous section hold in arbitrary dimension and for any form of the connection probability $p(\chi)$ and weight probability density $f(\epsilon)$ in Eqs. (1) and (3). To validate the theoretical calculations, we particularize to the $\mathbb{S}^1$ model as generator of the topology [1]. In that model, we choose the circle $\mathbb{S}^1$ of radius $R = N/2\pi$ to be the underlying geometry over which nodes are uniformly distributed with density $\delta = 1$. Distances among nodes are measured in terms of arc lengths, that is, two nodes with angular positions $\theta$ and $\theta'$ are therefore at a distance

$d(\theta, \theta') = R\Delta\theta$ where $\Delta\theta = \pi - |\pi - |\theta - \theta'||$. The connection probability is set to

$$p(\chi) = \frac{1}{1 + \chi^\beta} \quad \text{with} \quad \chi = \frac{d}{\mu\kappa\kappa'},$$ (30)

where $\beta > 1$ is a free parameter that can be used to tune the clustering. Equation (30) casts the ensemble of networks generated by the model into exponential random networks [5], i.e., networks that are maximally random given the constraints imposed by the free parameters (that is, $\rho(\kappa)$ and $\beta$). To obtain a scale-free degree distribution, hidden variables $\kappa$ are distributed according to

$$\rho(\kappa) = \frac{(\gamma - 1)\kappa_0^{\gamma-1}\kappa^{-\gamma}}{1 - (\kappa_c/\kappa_0)^{1-\gamma}}$$ (31)

with $\kappa_0 < \kappa < \kappa_c$ and $\gamma > 1$. Notice that by keeping the explicit dependence in the upper cut-off it is possible to model networks with $\gamma < 2$ and a hard cut-off, as found for instance in airports networks [6]. Moreover, since it is generally more convenient to fix the average degree $\langle k \rangle$ explicitly, we choose $\kappa_0$ such that

$$\langle\kappa\rangle = \int_{\kappa_0}^{\kappa_c} \kappa\rho(\kappa)d\kappa = \frac{(\gamma-1)\kappa_0}{(\gamma-2)}\frac{1 - (\kappa_c/\kappa_0)^{2-\gamma}}{1 - (\kappa_c/\kappa_0)^{1-\gamma}} = \langle k \rangle \ ,$$ (32)

and fix the remaining free parameters $\kappa_c$ and $\gamma$ externally. From Eq. (2), we expect

$$P(k) = \frac{(\gamma-1)\kappa_0^{\gamma-1}}{1 - (\kappa_c/\kappa_0)^{1-\gamma}}\frac{\Gamma(k - \gamma + 1, \kappa_0, \kappa_c)}{k!} \sim \frac{(\gamma-1)\kappa_0^{\gamma-1}k^{-\gamma}}{1 - (\kappa_c/\kappa_0)^{1-\gamma}} \sim k^{-\gamma} \ ,$$ (33)

where $\Gamma(x, \kappa_0, \kappa_c)$ is the generalized incomplete gamma function. It is defined as $\Gamma(t, z_0, z_1) = \int_{z_0}^{z_1} z^{t-1}e^{-z}dz$ (the regular complete and incomplete gamma functions can be retrieved by setting the bounds $z_0$ and $z_1$ to the appropriate values). Whenever $1 \sim z_0 \ll z_1$, a condition holding for most realistic degree distributions, the double incomplete gamma function scales as $\Gamma(n + \epsilon, \kappa_0, \kappa_c) \sim \Gamma(n)n^\epsilon$ with $n \in \mathbb{N}$ and $\epsilon \in \mathbb{R}$ [7]. Note also that it is common to set $\kappa_c = \kappa_0 N^{1/(\gamma-1)}$, i.e., the *natural* cut-off of a scale-free distribution [8, 9]. However, in general $\kappa_c$ can take any value in response to particular mechanisms at play, like the limited capacity to handle more than a given number of connections in the airports network.

To assign weights on top of the topology generated by the model, the noise distribution in Eq. (3) is chosen to be a gamma distribution of average $\langle \epsilon \rangle = 1$, that is,

$$f(\epsilon) = \frac{\lambda^\lambda}{\Gamma(\lambda)}\epsilon^{\lambda-1}e^{-\lambda\epsilon} \ \text{with} \ \langle\epsilon^2\rangle = 1 + \frac{1}{\lambda}.$$ (34)

This particular choice allows us to interpolate with a single parameter between a zero noise limit when $\lambda \gg 1$, exponential noise when $\lambda = 1$, and strongly heterogeneous noise when $\lambda \ll 1$. Finally, to control the correlation between strength and degree and, therefore, to tune the strength distribution, we assume a deterministic relation between hidden variables $\sigma$ and $\kappa$ of the form $\sigma = a\kappa^\eta$, as observed in real complex networks [6]. From Eq. (6), we thus expect

$$\bar{s}(k) = \frac{ak\Gamma(k - \gamma + \eta, \kappa_0, \kappa_c)}{\Gamma(k - \gamma + 1, \kappa_0, \kappa_c)} \sim ak^\eta \ .$$ (35)

Notice that the relation between average strength and degree in the previous expression is totally independent of the underlying metric space. It implies that the strength distribution scales as $P(s) \sim s^{-\xi}$ for $s \gg 1$ with $\xi = (\gamma + \eta - 1)/\eta$. Supplementary Figure 1 shows the basic topological and weighted properties of a network generated using Eqs. (30)–(35) and compares them to the theoretical predictions presented in this section. Apart from some expected fluctuations due to finite size, the agreement between the two is excellent.

## C. The effect of the underlying geometry

Geometry has a strong effect on the strength and weight distributions, which depend on the coupling parameter $\alpha$. In fact, as shown in Sec. I A 3, the second moment of the strength distribution $\langle s^2 \rangle$ is proportional to the integral $\int_0^\infty \chi^{D-2\alpha-1} p(\chi) d\chi$, which diverges whenever $\alpha > D/2$. The origin of these fluctuations is rooted in the strong constraints that geometry imposes on the weights of individual links. In the absence of coupling (i.e., $\alpha = 0.0$) the metric distance between nodes does not influence the magnitudes of the weights. Consequently, the distribution of weights generated by the model is the original pdf in Eqs. (3) and (4) convoluted with the distribution of values of the ratio $\sigma\sigma'/\kappa\kappa'$. Conversely, in the case of strong coupling (i.e., $\alpha \lesssim 1$), short range links are constrained to have larger weights whereas long range ones have small weights. This effect increases the heterogeneity in the weight distribution and causes the divergence of $\langle s^2 \rangle$ when $\alpha > D/2$. Supplementary Figure 2a shows this effect on synthetic networks generated with the model with identical parameters except for the value of $\alpha$.

The same effect is visible in the local heterogeneity of the weights attached to a given node. To characterize such heterogeneity, we use the disparity measure defined as

$$Y_i = \sum_j \left( \frac{w_{ij}}{s_i} \right)^2 , \qquad (36)$$

where $w_{ij}$ is the weight of the link between nodes $i$ and $j$ ($w_{ij} = 0$ if there is no link) and $s_i = \sum_j w_{ij}$ [10]. In Supplementary Figure 2b, we see that in the absence of coupling (i.e., $\alpha = 0.0$) the disparity scales as $Y_i \sim k_i^{-1}$ corresponding to the situation in which weights are roughly homogeneously distributed among the links [11]. On the other side of the spectrum, we see that under maximal coupling (i.e., $\alpha \lesssim 1$), the disparity decreases slower than $k_i^{-1}$ meaning that weights are heterogeneous and that the large strength of nodes is due to a handful of links with large weights.

## II. TEST OF THE TRIANGLE INEQUALITY

To test the triangle inequality in a given weighted complex network, we first find the parameters $\mu, \beta$, and $\gamma$ that best match the empirical topology. To achieve the optimal matching, we use the empiric sequence of degrees as input for the sequence of $\kappa$'s so that the fluctuations in the tail of the degree distribution of the input network are preserved. The sequence is then used to generate different weighted networks as follows. From the empiric relation strength-degree, we measure the proportionality factor $a$ and the exponent $\eta$, as well as the first and second moments of the strength distribution $\langle s \rangle$ and $\langle s^2 \rangle$. For fixed values of $\alpha$ and of the fluctuations of the pdf function $f(\cdot)$, $I_4 = \langle \epsilon^2 \rangle$, we generate a large number of synthetic weighted networks and measure the average value of $CV^2(s) = \text{Var}[s^2]/\langle s \rangle^2$ and its ensemble fluctuations. From Sec. I A 3, we expect the average value of $CV^2(s)$ to scale linearly with $\langle \epsilon^2 \rangle$ as

$$CV^2(s) = \frac{\langle s^2 \rangle}{\langle s \rangle^2} - 1 = \frac{\Gamma(D/2)\langle \sigma^2/\kappa \rangle^2 I_5}{2\pi^{D/2}\delta\mu\langle\sigma\rangle^4 I_2^2 I_3^2}\langle \epsilon^2 \rangle . \qquad (37)$$

The left hand side in this equation can be directly measured from the network. The right hand side depends linearly on the noise $\langle \epsilon^2 \rangle$ whereas its pre-factor depends both on the topology and on $\alpha_{\text{real}}$ through the integrals $I_3$ and $I_5$ (see Sec. I for technical details).

In Supplementary Figure 3a, we show $CV^2(s)$ as a function of $\langle \epsilon^2 \rangle$ and different values of $\alpha$ for one of the synthetic networks used in Supplementary Figure 4 with $\alpha_{\text{origin}} = 0.4$ and noise

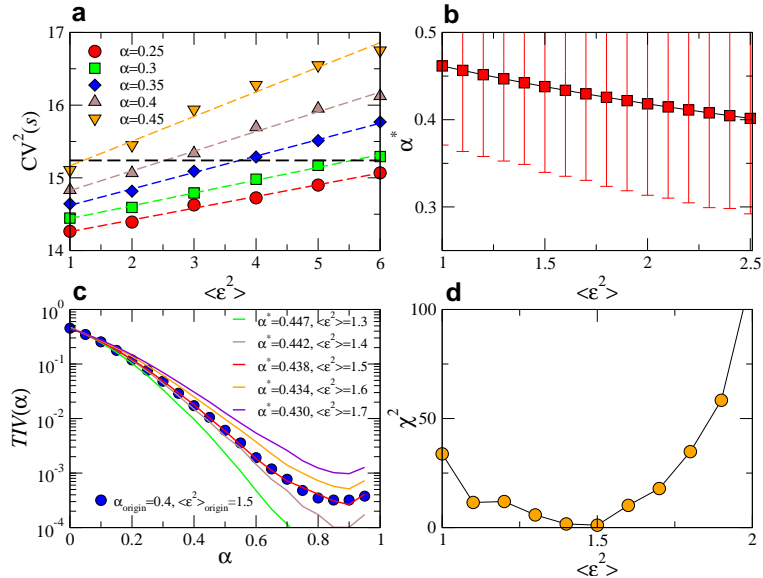

SUPPLEMENTARY FIGURE 3. **Illustration of the test of the triangle inequality** Test of the triangle inequality applied to a synthetic network generated with $\alpha_{\text{origin}} = 0.4$ and noise $\langle \epsilon^2 \rangle_{\text{origin}} = 1.5$. **a**, the square of the coefficient of variation of nodes' strength as a function of the noise $\langle \epsilon^2 \rangle$ in synthetic weighted networks with different values of $\alpha$. The horizontal dashed line is the empirical value measured in the input network. **b**, values of $\alpha^*$ as a function of the noise obtained from the intersection of the dashed line in **a** with the synthetic curves. **c**, $TIV(\alpha)$ curves for synthetic networks with the values of $\alpha^*$ and $\langle \epsilon^2 \rangle$ from **b** compared to the same function for the input network. **d**, $\chi^2$ statistics obtained from the comparison of function $TIV(\alpha)$ between the model and input network.

$\langle \epsilon^2 \rangle_{\text{origin}} = 1.5$. The intersection of these curves with the empirical value of $CV^2(s)$ defines a collection of $\alpha$'s as a function of the noise, $\alpha^* \left( \langle \epsilon^2 \rangle \right)$, which become the potential candidates to be the estimate of $\alpha_{\text{real}}$ (see Supplementary Figure 3**b**). Finally, for each pair $\left( \langle \epsilon^2 \rangle, \alpha^* \right)$ in Supplementary Figure 3**b** we measure the function $TIV(\alpha)$ and compare it with the same function measured in the input network (see Supplementary Figure 3**c**). The comparison is performed by measuring the standard $\chi^2$ statistics. The inferred value of $\alpha_{\text{real}}$ corresponds to the value of $\alpha^*(\langle \epsilon^2 \rangle)$ minimizing the value of $\chi^2$ (see Supplementary Figure 3**d**). To find a lower bound of the inferred value of $\alpha_{\text{real}}$, we use the ensemble fluctuations of $CV^2(s)$. For any fixed value of $\langle \epsilon^2 \rangle$, the lower bound of $\alpha^*$ is the value of $\alpha$ that is still able to reproduce the empirical value of $CV^2(s)$ (see Supplementary Figure 3**b**). The lower bound for the inferred value of $\alpha_{\text{real}}$ is the lower bound of $\alpha^*$ that corresponds to the optimal value of the noise $\langle \epsilon^2 \rangle$ in Supplementary Figure 3**d**. Supplementary Figure 4 shows the result of this method in the case of synthetic networks with different values of $\alpha$ and noise levels. As it can be seen, the inferred values of $\alpha$ match the true values in most of the cases. In fact, these fluctuations are due to the fact that the noise measured in the original synthetic network does not necessarily equal the typical noise of the ensemble of networks generated using the same parameters. Our method also allows us to find lower bounds on the inferred values, as shown by the grey areas in Supplementary Figure 4. The most remarkable aspect of the test is that it can be performed without any explicit embedding of the network and, thus, it can be readily applied to real networks for which an embedding is not available.

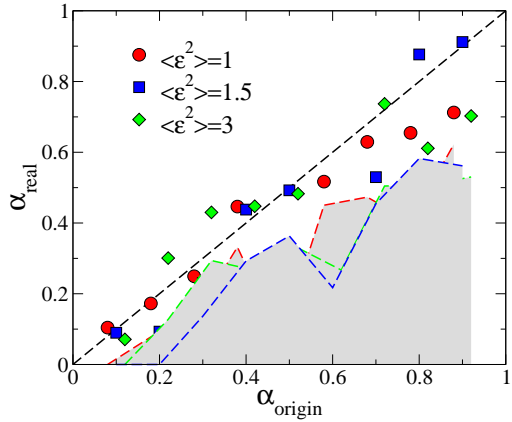

SUPPLEMENTARY FIGURE 4. **Validation of the test of the triangle inequality** $\alpha_{\text{real}}$ vs. $\alpha_{\text{origin}}$ for synthetic networks generated with the model for different values of noise $\langle \epsilon^2 \rangle$. In all cases, network topologies are generated with $\gamma = 2.5$, $\beta = 2$, $\eta = 1.2$, $\langle k \rangle = 10$, and $N = 10^4$. The solid grey area indicates the lower bounds found by the method.

## A. Violation of the triangle inequality

We expect the violation of the triangle inequality to depend essentially on the level of noise in the system $\langle \epsilon^2 \rangle$ through the term in the right hand side of Eq. (7) in the main text. To a lesser extent, the violation may also be due to the fact that the hidden variables $\kappa$ and $\sigma$ are approximated by the actual degree and the strength, respectively, of nodes. For most of the analysed real networks, the percentage of violations is very small (of the order of few percent) whereas in the case of the cargo ships network it is close to 20%, due to the high level of noise present in the system. In short, our model predicts that there should not be any dependence on the degree in the nodes belonging to triangles that violate the triangle inequality. To test this prediction, we have measured explicitly the average degree of such nodes as compared to the average degree of nodes in all triangles (see Supplementary Figure 5). In many cases the average degree is very similar, thus confirming our prediction. The largest discrepancy is found in the metabolic network. However, notice that this network has a very small percentage of violations, which makes it more prone to statistical fluctuations.

## B. Behaviour of $TIV(\alpha)$ with $\alpha \sim 1$

The increase of $TIV(\alpha)$ close to $\alpha = 1$ on Supplementary Figure 3 and on Figs. **3a–b** in the main text is expected and is in fact an artefact of Eq. (5) and of our our choice of the probability of connection [i.e., Eq. (30)]. Indeed, substituting Eq. (23) in Eq. (7) in the main text and neglecting the noise term (whose mean value is close to zero) we obtain

$$\ln \left[ \frac{\omega_{ij}\omega_{jk}}{\omega_{ik}} \left( \frac{\kappa_j}{\sigma_j} \right)^2 \right] \leq \frac{R}{2}\alpha + \ln \left( \sin \left[ \frac{(1-\alpha)\pi}{\beta} \right] \right) + \ln \left( \frac{\beta}{2\pi\mu\langle\sigma\rangle} \right) \ . \tag{38}$$

Supplementary Figure 6 shows the behaviour of $\alpha$-dependent terms of the right hand side of Eq. (38) for the real networks considered in the main text. For the low values of $\alpha$, we see that the right hand side of Eq. (38) is an increasing function which implies that $TIV(\alpha)$ decreases with increasing $\alpha$ (i.e., it is *more and more difficult* to violate the triangle inequality as $\alpha$ increases). However, all curves reach a plateau at $\alpha \simeq 0.8$ after which they start to decrease. As expected, these plateaus

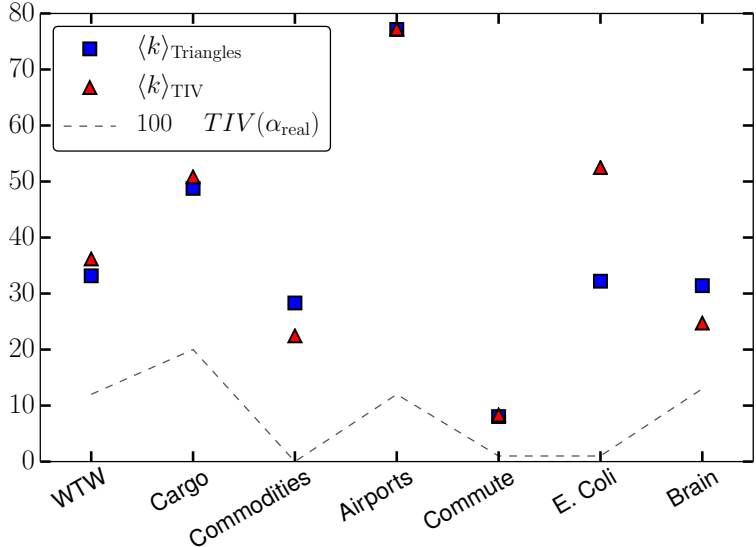

SUPPLEMENTARY FIGURE 5. **Further properties of triangles** Average degree of nodes in triangles that violate the triangle inequality ($\langle k \rangle_{\mathrm{TIV}}$) and in all triangles ($\langle k \rangle_{\mathrm{Triangle}}$) for the networks considered in the main text. The average is performed by sampling over triangles which implies that the degree of a node is weighted by the number of triangles to which it participates (as in Fig. 1 of the main text). The dashed line shows the fraction of triangles that violate the triangle inequality when using the inferred value $\alpha_{\mathrm{real}}$.

correspond to the points where the $TIV(\alpha)$ start to increase (for some networks this increase is not visible due to the linear scale of the y axis).

### C.  Application to real networks

We applied this methodology to the real networks mentioned in the paper and Supplementary Table 1 shows the parameters thus inferred (see also Fig. 3 in the main text). The comparison between the properties of networks generated using these parameters with the ones of the original real networks is shown on Supplementary Figures 7–12. Besides some expected fluctuations inherent to the model (i.e., only one synthetic network is used for each figure), these figures confirm that the model can reproduce many topological and weighted features observed in real complex networks.

### III.  COMPARISON WITH OTHER MODELS

We present further evidence to support the claim that our model is the most accurate approach to model real weighted complex networks. To do so, we show the results obtained by using the two models introduced in Refs. [6] and [12], as well as a new one that generalizes them. These models use the original network topology randomized under the constraints of preserving the degree sequence and the average clustering coefficient using the software developed in Refs. [13, 14]. Weights are then assigned to each link according to the following rules:

- **model A:** $w_{ij} \propto (k_i k_j)^\theta$, where $k_i$ and $k_j$ are the degrees of nodes $i$ and $j$, respectively;

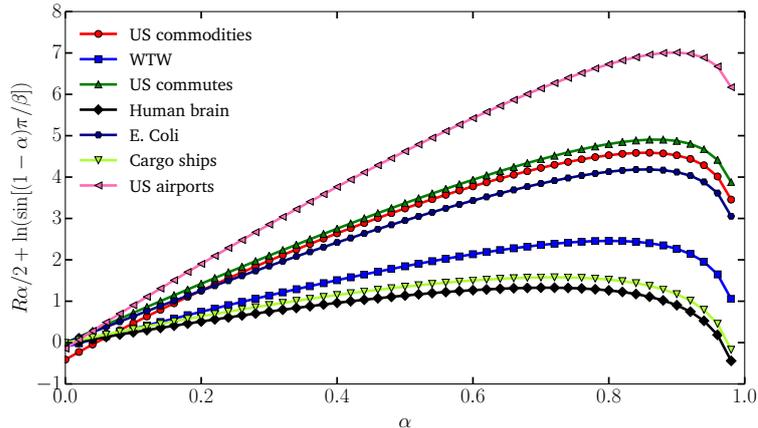

SUPPLEMENTARY FIGURE 6. **Behaviour of the violation threshold** $\alpha$-dependent terms of the right hand side of Eq. (38) as a function of $\alpha$ for the real networks considered in the main text.

- **model B:** $w_{ij} \propto (c_i c_j)^\delta$, where $c_i$ and $c_j$ are the clustering coefficient of nodes $i$ and $j$, respectively;

- **model C:** $w_{ij} \propto (k_i k_j)^\mu (c_i c_j)^\nu$. This model accounts for the fact that weights among high degree nodes are higher but also that weights among highly clustered nodes are also higher.

The exponents $\theta$, $\delta$, $\mu$ and $\nu$ are chosen as those minimizing the $\chi^2$ statistic for the corresponding dataset (see the captions of Supplementary Figures 14–41 for the inferred values).

Although the three models reproduce exactly the degree sequence, the degree-dependent clustering of the randomized networks is never better than the one obtained with our model. We find that models A and C can reproduce fairly well the strength distributions, or at least their general shapes. This is due to the strong influence of the topology over the weighted organization, and it illustrates well the reason why we factorized the weights on Fig. 1 in the main text to account for the effect of the topology. However, except for the world trade web and US airports network, we find that the three models reproduce poorly the weight distributions and the disparities. This is not particularly surprising in the case of the US airports network since our model predicts a weaker dependence on the metric space, leaving weights mainly as a function of the degree of nodes. Similarly, it is not surprising in the case of the world trade web given its small size. Nevertheless, even though some of the local properties can be reproduced in some of the networks by the three models, Supplementary Figures 14–41 show that none of them can reproduce the $TIV(\alpha)$ curves observed for the real networks, suggesting that our assumption about the metric origin of weights is a much better explanation of the real data.

| Name | $\rho_{m,\omega}$ | $\rho_{m,\omega^{\mathrm{norm}}}$ | $\beta$ | $\alpha$ | $\langle \epsilon^2 \rangle$ | $\eta$ | $a$ | Suppl. Figure |
|---|---|---|---|---|---|---|---|---|
| World Trade | 0.68 | 0.10 | 2.5 | 0.41 | 1.3 | 1.63 | 3772 | 7 |
| Cargo ships | 0.19 | 0.10 | 1.85 | 0.65 | 1.7 | 1.05 | 83 | 8 |
| US Commodities | 0.24 | 0.05 | 1.3 | 0.65 | 1.2 | 1.22 | 3045 | 9 |
| US Airports | 0.72 | 0.03 | 1.4 | 0.15 | 1.4 | 1.72 | 10000 | 10 |
| US Commute | 0.51 | 0.17 | 2.2 | 0.59 | 1.4 | 2.02 | 719 | 11 |
| E. Coli | 0.73 | 0.45 | 2.2 | 0.45 | 1.3 | 1.09 | 1 | 12 |
| Human brain | 0.27 | 0.23 | 2.8 | 0.45 | 1.3 | 0.86 | 0.015 | 13 |

SUPPLEMENTARY TABLE 1. **Parameters and information about the datasets** Pearson correlation coefficients between the multiplicity, $m$, and the weight and normalized weight of links in the real networks considered in the paper. Also, the parameters used to reproduce real networks with our model (see the main text for description).

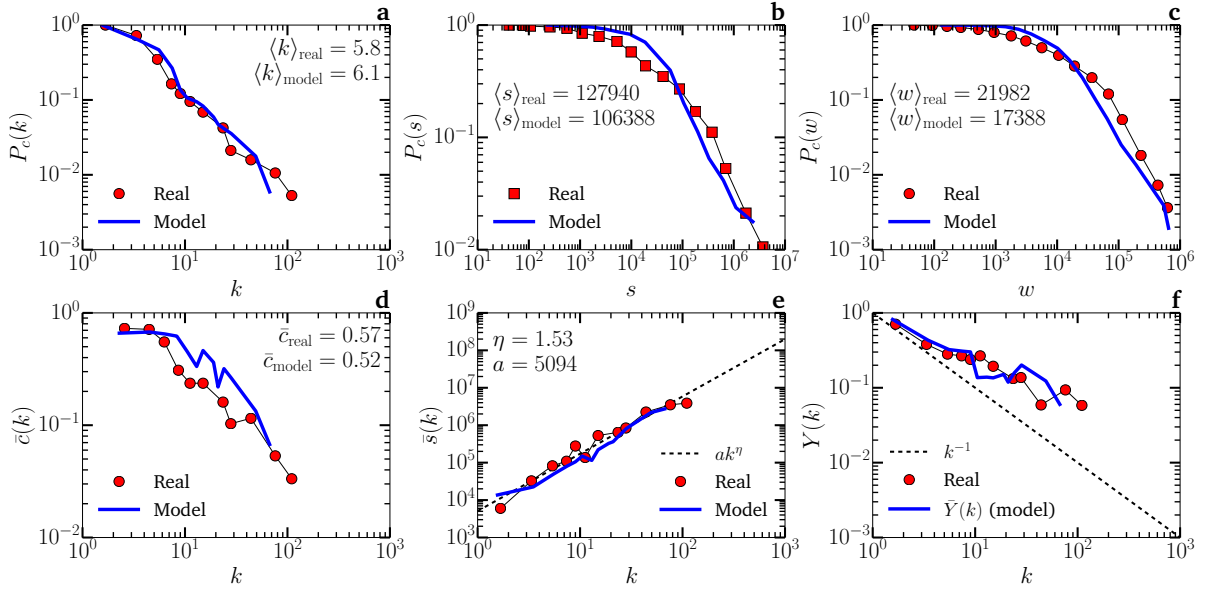

SUPPLEMENTARY FIGURE 7. **Predictions by the model introduced in Sec. I.** Comparison between topological and weighted properties of the WTW (symbols) and a synthetic network generated by the model with the parameters given in Supplementary Table 1 (solid lines). **a**, complementary cumulative degree distribution. **b**, complementary cumulative strength distribution. **c**, complementary cumulative weight distribution of links. **d**, degree-dependent clustering coefficient. **e**, average strength of nodes of degree $k$. **f**, disparity of nodes as a function of their degree.

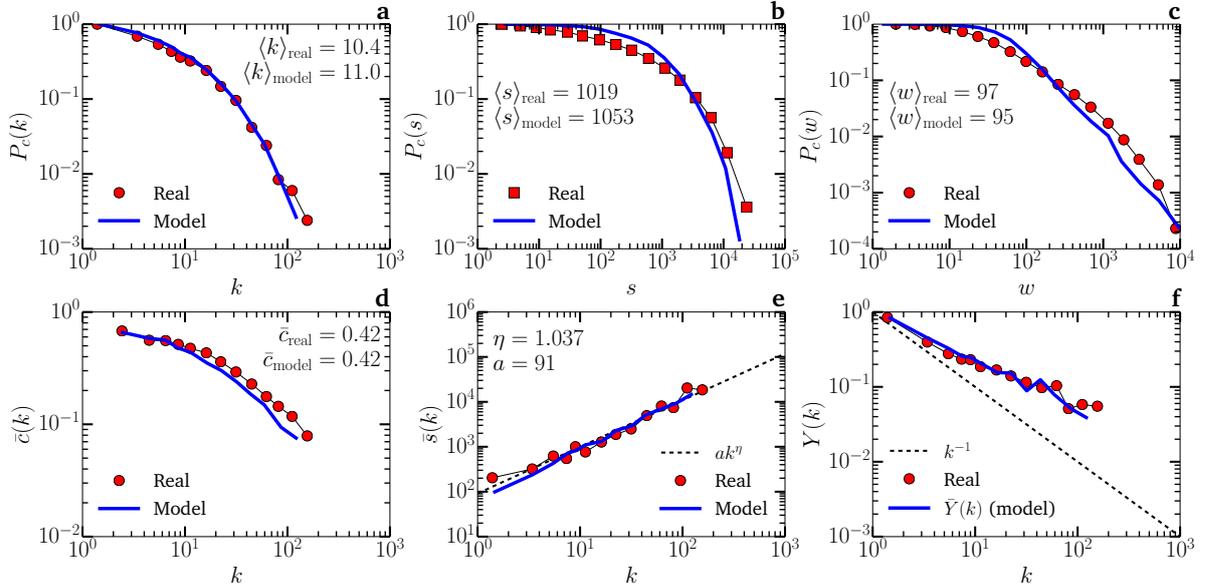

SUPPLEMENTARY FIGURE 8. **Predictions by the model introduced in Sec. I.** Comparison between topological and weighted properties of the Cargo ships network (symbols) and a synthetic network generated by the model with the parameters given in Supplementary Table 1 (solid lines). **a**, complementary cumulative degree distribution. **b**, complementary cumulative strength distribution. **c**, complementary cumulative weight distribution of links. **d**, degree-dependent clustering coefficient. **e**, average strength of nodes of degree $k$. **f**, disparity of nodes as a function of their degree.

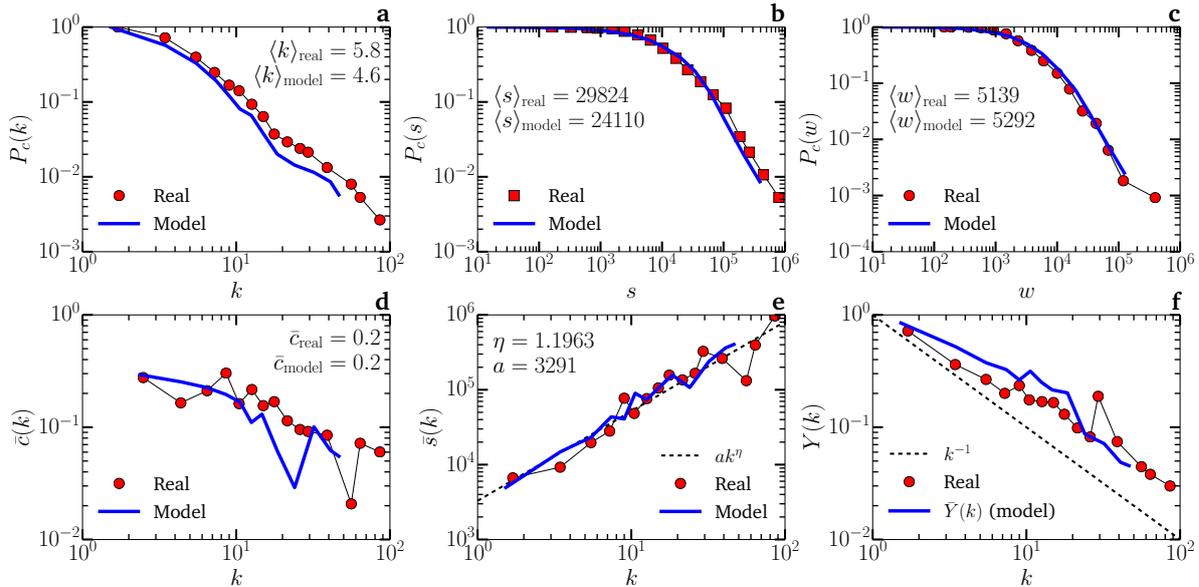

SUPPLEMENTARY FIGURE 9. **Predictions by the model introduced in Sec. I.** Comparison between topological and weighted properties of the US Commodities network (symbols) and a synthetic network generated by the model with the parameters given in Supplementary Table 1 (solid lines). **a**, complementary cumulative degree distribution. **b**, complementary cumulative strength distribution. **c**, complementary cumulative weight distribution of links. **d**, degree-dependent clustering coefficient. **e**, average strength of nodes of degree $k$. **f**, disparity of nodes as a function of their degree.

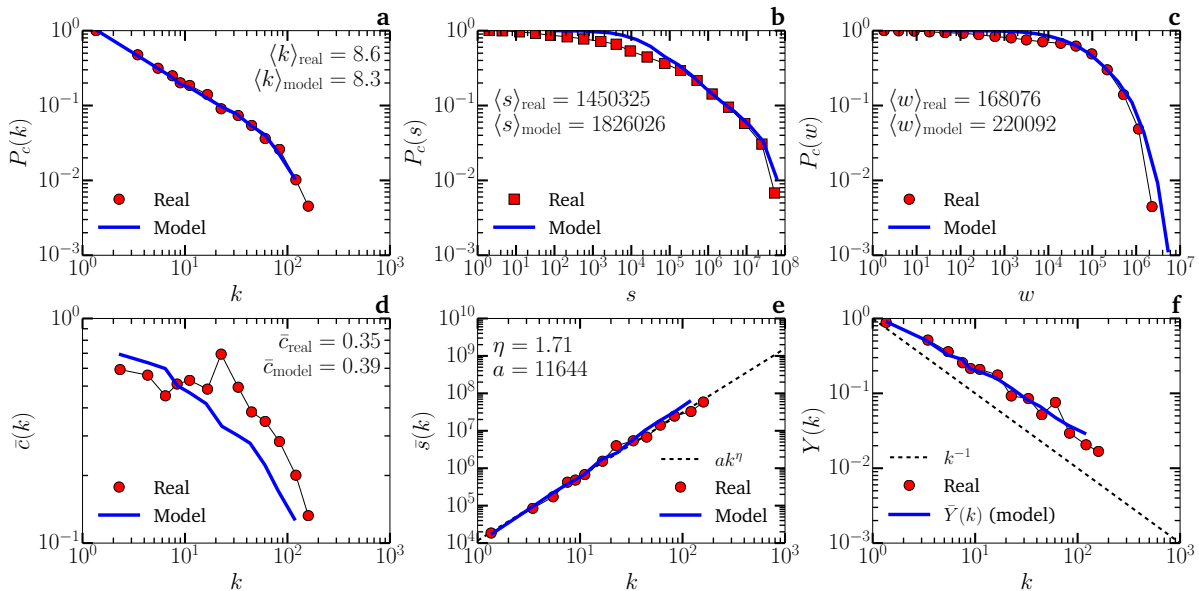

SUPPLEMENTARY FIGURE 10. **Predictions by the model introduced in Sec. I.** Comparison between topological and weighted properties of the US airports network (symbols) and a synthetic network generated by the model with the parameters given in Supplementary Table 1 (solid lines). **a**, complementary cumulative degree distribution. **b**, complementary cumulative strength distribution. **c**, complementary cumulative weight distribution of links. **d**, degree-dependent clustering coefficient. **e**, average strength of nodes of degree $k$. **f**, disparity of nodes as a function of their degree.

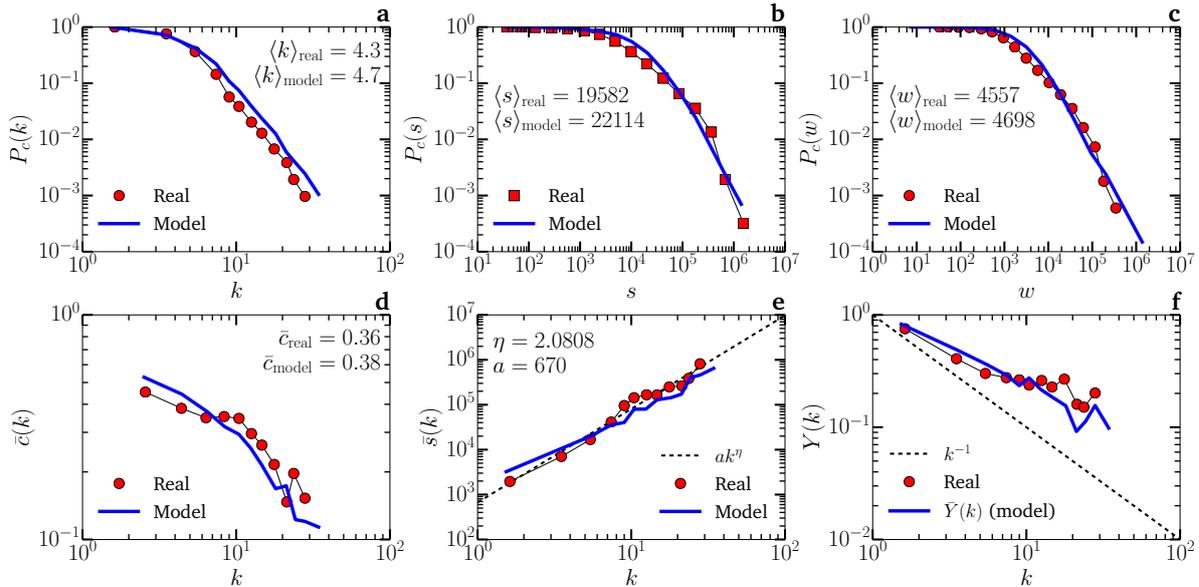

SUPPLEMENTARY FIGURE 11. **Predictions by the model introduced in Sec. I.** Comparison between topological and weighted properties of the US Commute network (symbols) and a synthetic network generated by the model with the parameters given in Supplementary Table 1 (solid lines). **a**, complementary cumulative degree distribution. **b**, complementary cumulative strength distribution. **c**, complementary cumulative weight distribution of links. **d**, degree-dependent clustering coefficient. **e**, average strength of nodes of degree $k$. **f**, disparity of nodes as a function of their degree.

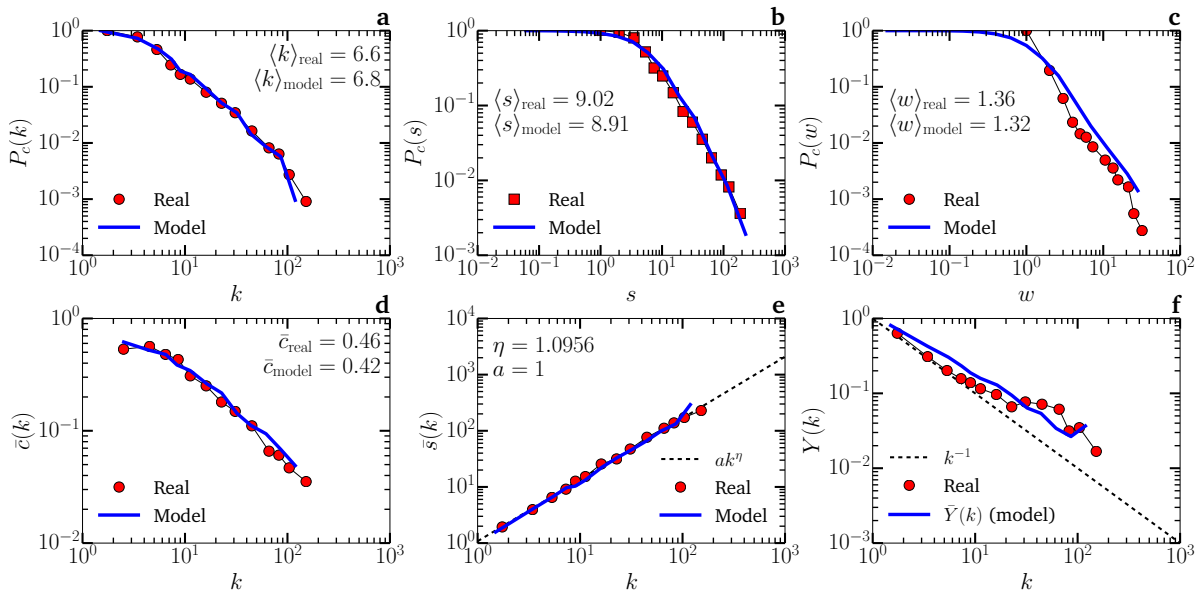

SUPPLEMENTARY FIGURE 12. **Predictions by the model introduced in Sec. I.** Comparison between topological and weighted properties of the iJO1366 *E. Coli* metabolic network (symbols) and a synthetic network generated by the model with the parameters given in Supplementary Table 1 (solid lines). **a**, complementary cumulative degree distribution. **b**, complementary cumulative strength distribution. **c**, complementary cumulative weight distribution of links. **d**, degree-dependent clustering coefficient. **e**, average strength of nodes of degree $k$. **f**, disparity of nodes as a function of their degree.

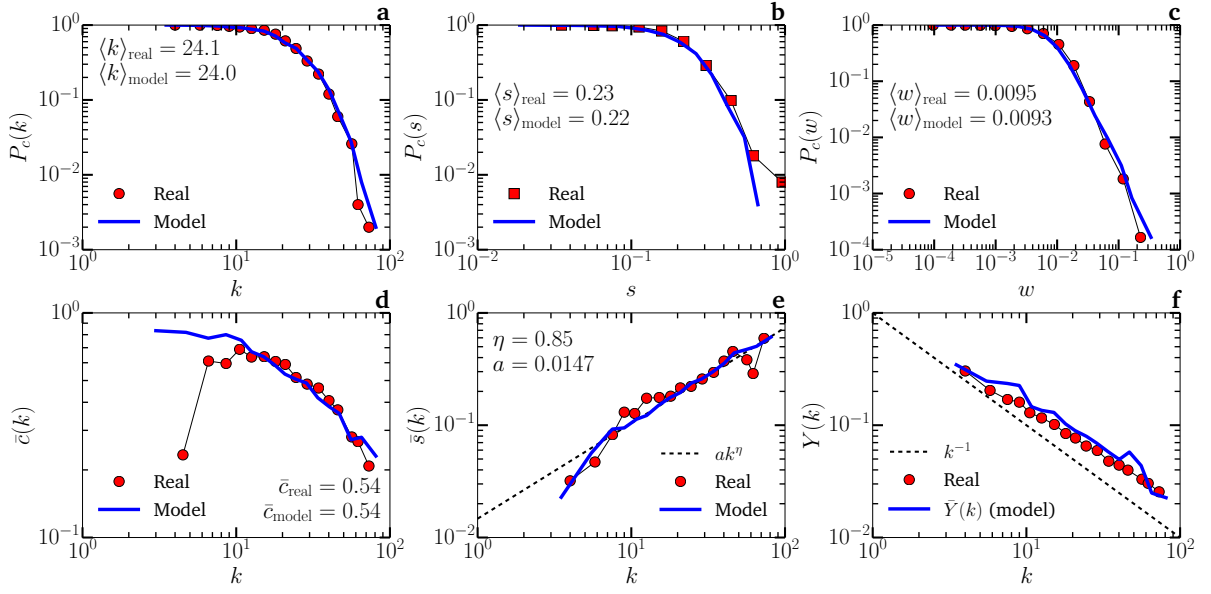

SUPPLEMENTARY FIGURE 13. **Predictions by the model introduced in Sec. I.** Comparison between topological and weighted properties of the Human brain network (symbols) and a synthetic network generated by the model with the parameters given in Supplementary Table 1 (solid lines). **a**, complementary cumulative degree distribution. **b**, complementary cumulative strength distribution. **c**, complementary cumulative weight distribution of links. **d**, degree-dependent clustering coefficient. **e**, average strength of nodes of degree $k$. **f**, disparity of nodes as a function of their degree.

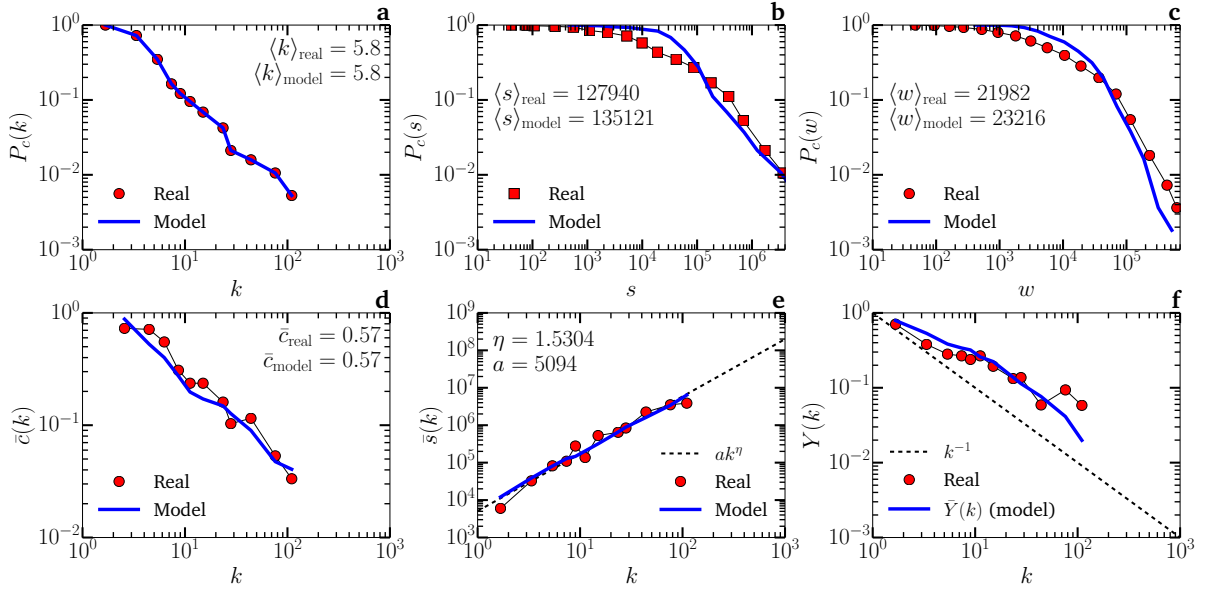

SUPPLEMENTARY FIGURE 14. **Predictions by Model A.** Comparison between topological and weighted properties of the WTW (symbols) and a synthetic network generated by model A with $\theta = 0.9$ (solid lines). **a**, complementary cumulative degree distribution. **b**, complementary cumulative strength distribution. **c**, complementary cumulative weight distribution of links. **d**, degree-dependent clustering coefficient. **e**, average strength of nodes of degree $k$. **f**, disparity of nodes as a function of their degree.

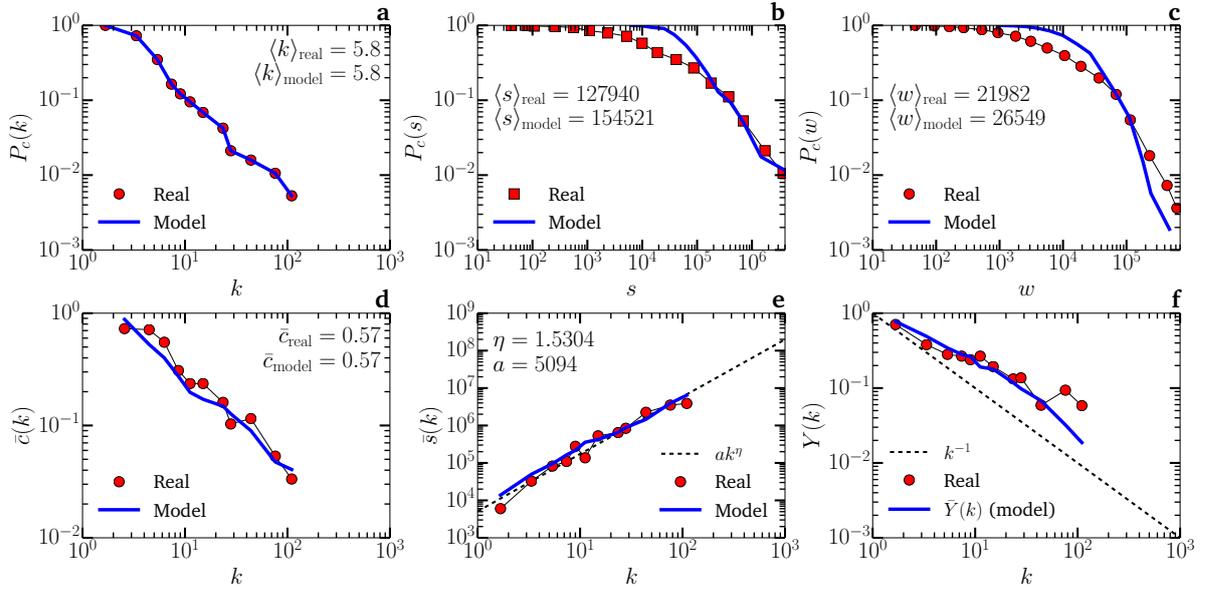

SUPPLEMENTARY FIGURE 15. **Predictions by Model B.** Comparison between topological and weighted properties of the WTW (symbols) and a synthetic network generated by model B with $\delta = -1.01$ (solid lines). **a**, complementary cumulative degree distribution. **b**, complementary cumulative strength distribution. **c**, complementary cumulative weight distribution of links. **d**, degree-dependent clustering coefficient. **e**, average strength of nodes of degree $k$. **f**, disparity of nodes as a function of their degree.

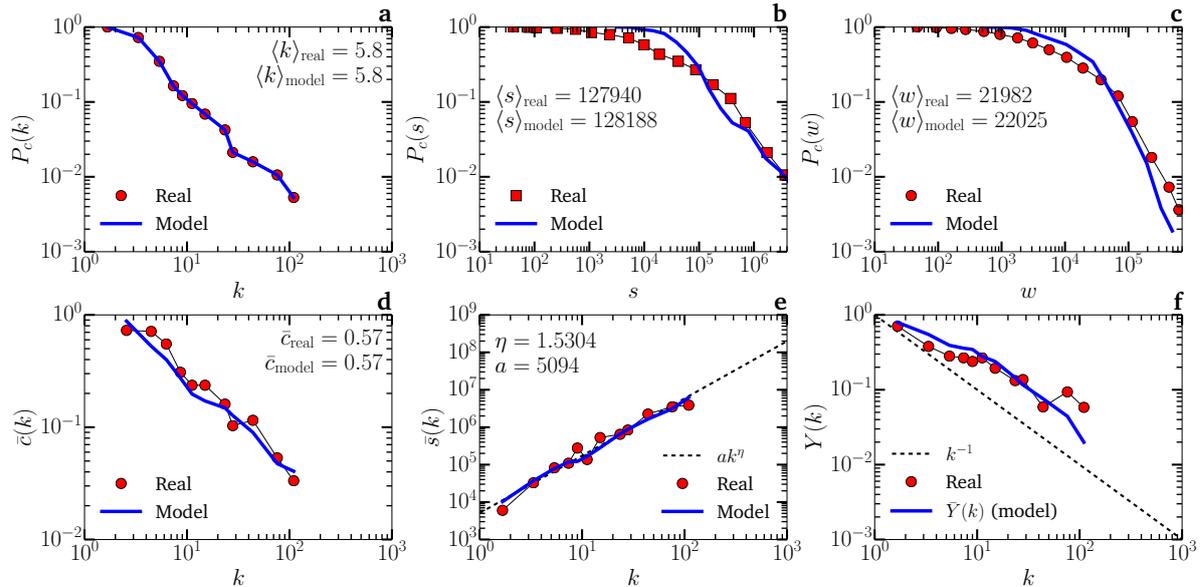

SUPPLEMENTARY FIGURE 16. **Predictions by Model C.** Comparison between topological and weighted properties of the WTW (symbols) and a synthetic network generated by model C with $\mu = 1.3$ and $\nu = 0.475$ (solid lines). **a**, complementary cumulative degree distribution. **b**, complementary cumulative strength distribution. **c**, complementary cumulative weight distribution of links. **d**, degree-dependent clustering coefficient. **e**, average strength of nodes of degree $k$. **f**, disparity of nodes as a function of their degree.

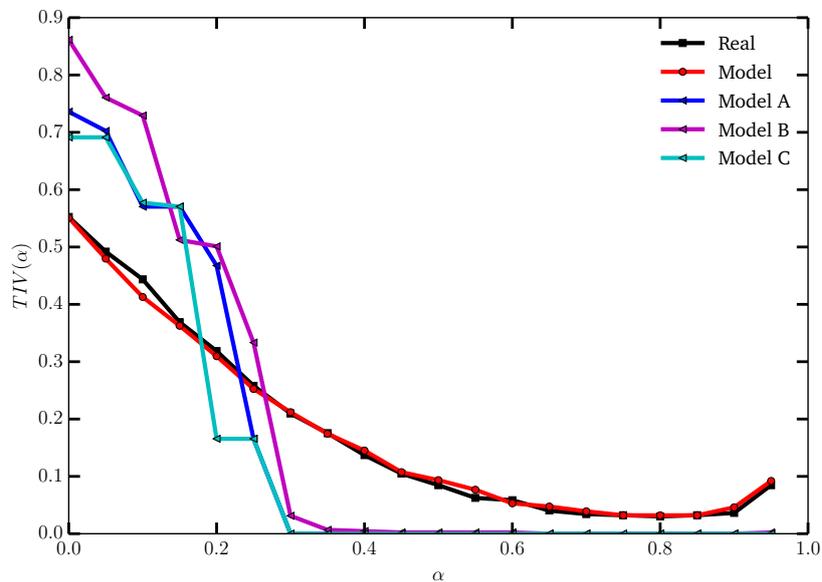

SUPPLEMENTARY FIGURE 17. **Triangle inequality violation spectrum.** Comparison between $TIV(\alpha)$ curves measured for the real WTW, our model and the models A, B and C with the exponents given in the caption of Supplementary Figures 14–16.

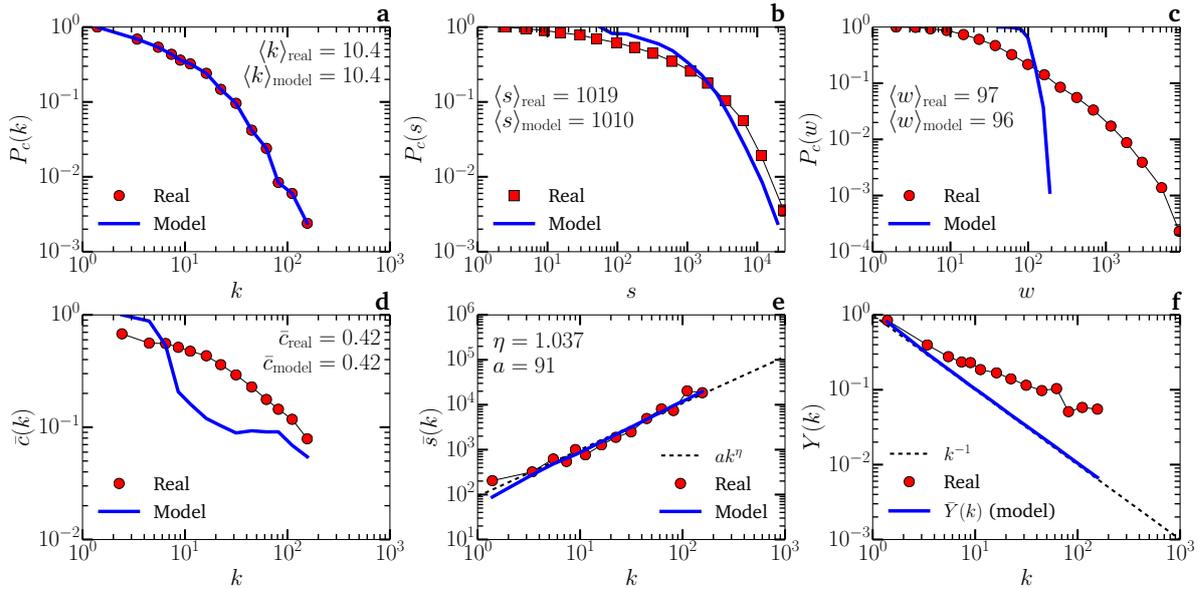

SUPPLEMENTARY FIGURE 18. **Predictions by Model A.** Comparison between topological and weighted properties of the Cargo ships network (symbols) and a synthetic network generated by model A with $\theta = 0.18$ (solid lines). **a**, complementary cumulative degree distribution. **b**, complementary cumulative strength distribution. **c**, complementary cumulative weight distribution of links. **d**, degree-dependent clustering coefficient. **e**, average strength of nodes of degree $k$. **f**, disparity of nodes as a function of their degree.

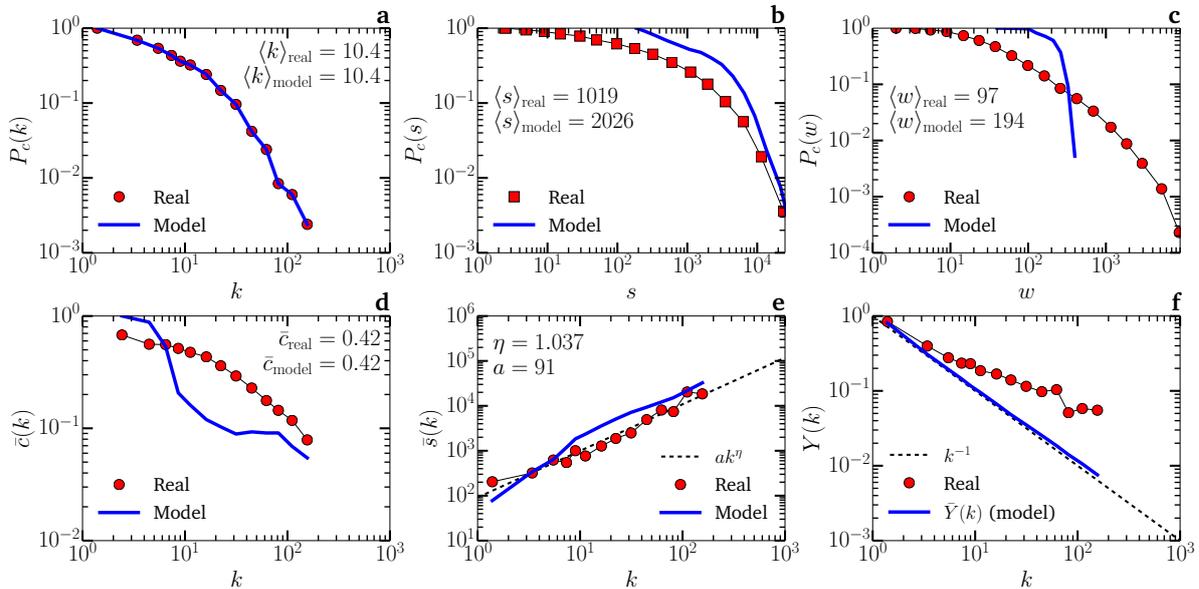

SUPPLEMENTARY FIGURE 19. **Predictions by Model B.** Comparison between topological and weighted properties of the Cargo ships network (symbols) and a synthetic network generated by model B with $\delta = -0.39$ (solid lines). **a**, complementary cumulative degree distribution. **b**, complementary cumulative strength distribution. **c**, complementary cumulative weight distribution of links. **d**, degree-dependent clustering coefficient. **e**, average strength of nodes of degree $k$. **f**, disparity of nodes as a function of their degree.

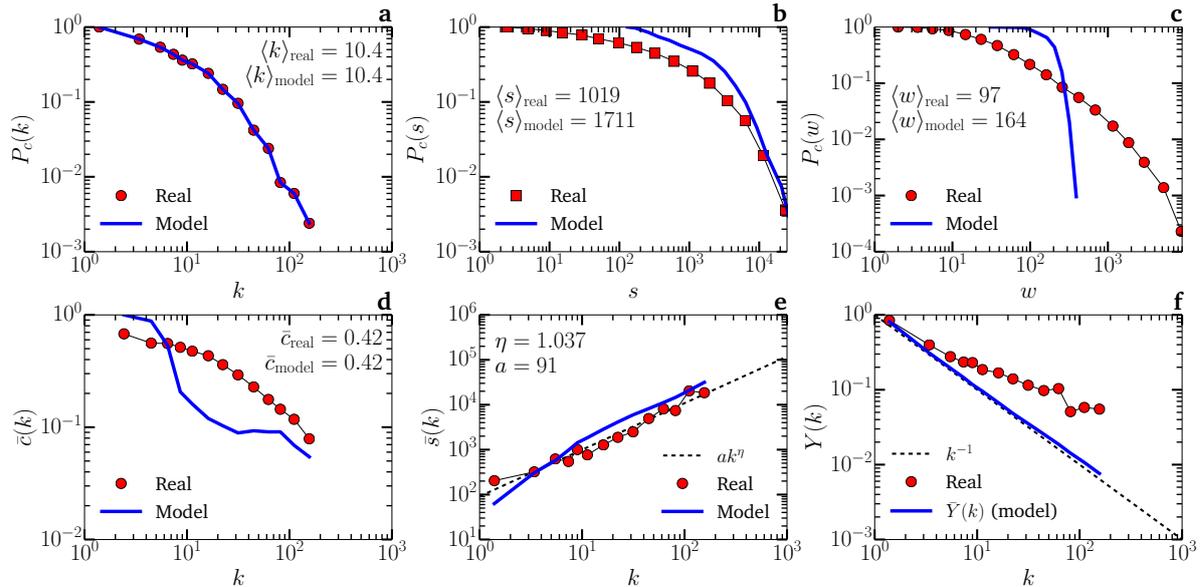

SUPPLEMENTARY FIGURE 20. **Predictions by Model C.** Comparison between topological and weighted properties of the Cargo ships network (symbols) and a synthetic network generated by model C with $\mu = 1.0$ and $\nu = -0.3$ (solid lines). **a**, complementary cumulative degree distribution. **b**, complementary cumulative strength distribution. **c**, complementary cumulative weight distribution of links. **d**, degree-dependent clustering coefficient. **e**, average strength of nodes of degree $k$. **f**, disparity of nodes as a function of their degree.

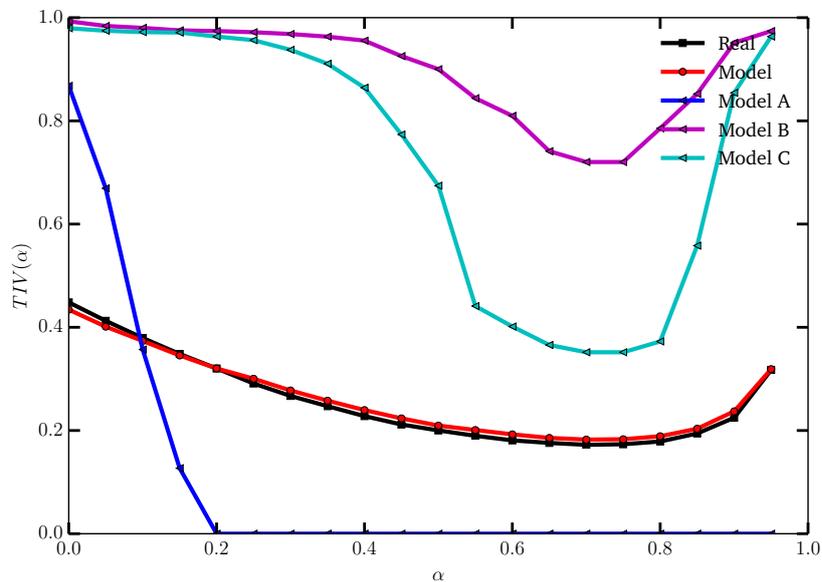

SUPPLEMENTARY FIGURE 21. **Triangle inequality violation spectrum.** Comparison between $TIV(\alpha)$ curves measured for the real Cargo ships network, our model and the models A, B and C with the exponents given in the caption of Supplementary Figures 18–20.

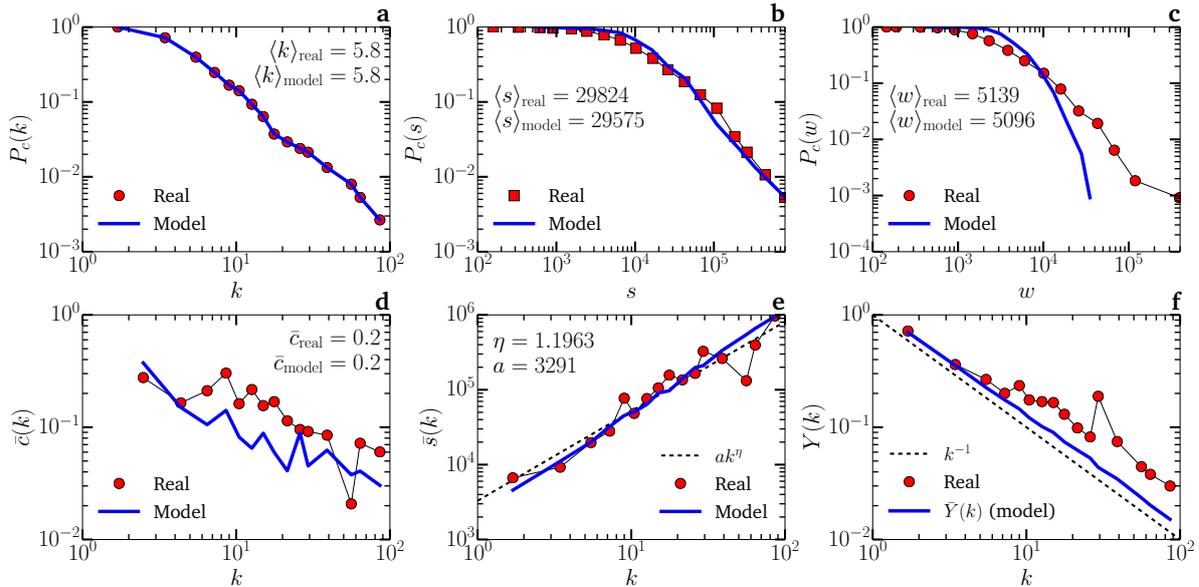

SUPPLEMENTARY FIGURE 22. **Predictions by Model A.** Comparison between topological and weighted properties of the US Commodities network (symbols) and a synthetic network generated by model A with $\theta = 0.51$ (solid lines). **a**, complementary cumulative degree distribution. **b**, complementary cumulative strength distribution. **c**, complementary cumulative weight distribution of links. **d**, degree-dependent clustering coefficient. **e**, average strength of nodes of degree $k$. **f**, disparity of nodes as a function of their degree.

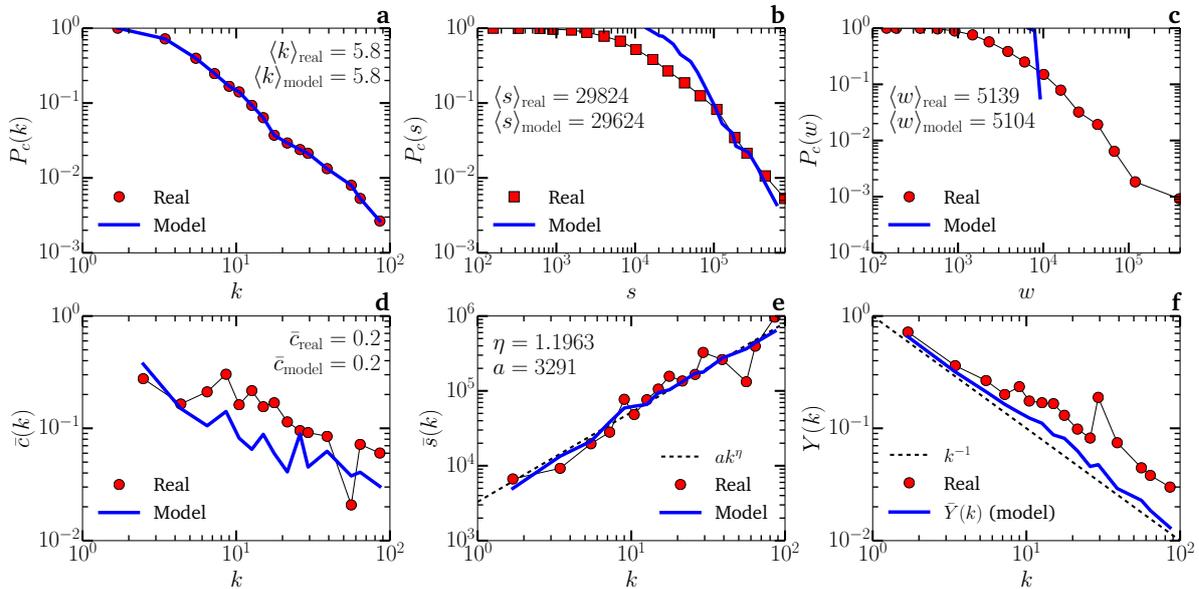

SUPPLEMENTARY FIGURE 23. **Predictions by Model B.** Comparison between topological and weighted properties of the US Commodities network (symbols) and a synthetic network generated by model B with $\delta = -0.07$ (solid lines). **a**, complementary cumulative degree distribution. **b**, complementary cumulative strength distribution. **c**, complementary cumulative weight distribution of links. **d**, degree-dependent clustering coefficient. **e**, average strength of nodes of degree $k$. **f**, disparity of nodes as a function of their degree.

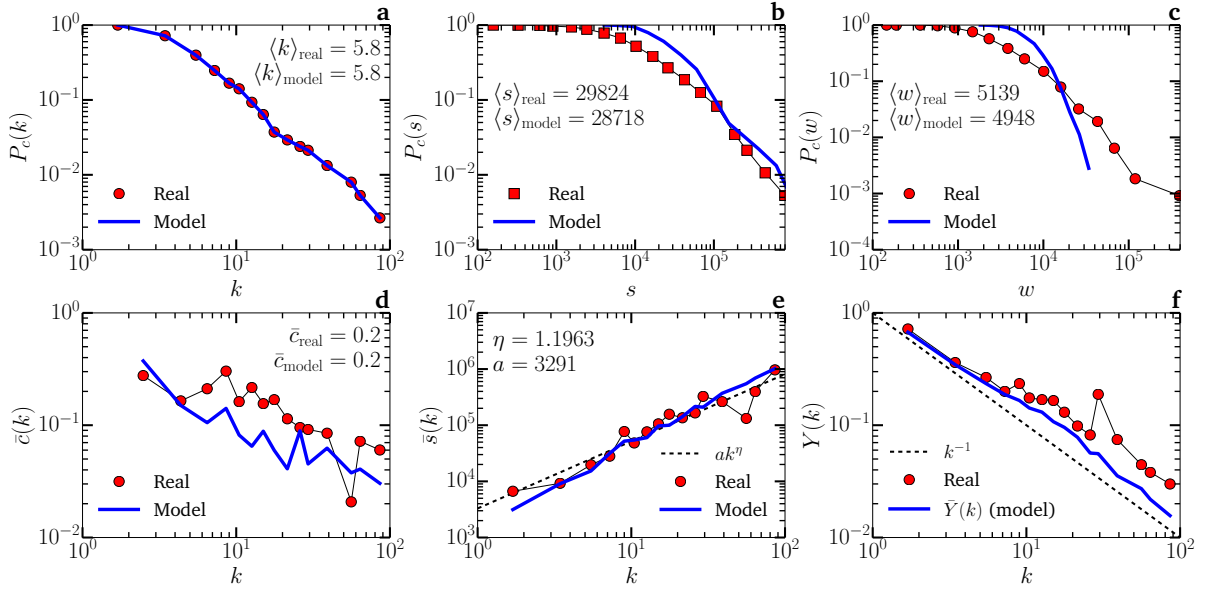

SUPPLEMENTARY FIGURE 24. **Predictions by Model C.** Comparison between topological and weighted properties of the US Commodities network (symbols) and a synthetic network generated by model C with $\mu = 0.425$ and $\nu = -0.025$ (solid lines). **a**, complementary cumulative degree distribution. **b**, complementary cumulative strength distribution. **c**, complementary cumulative weight distribution of links. **d**, degree-dependent clustering coefficient. **e**, average strength of nodes of degree $k$. **f**, disparity of nodes as a function of their degree.

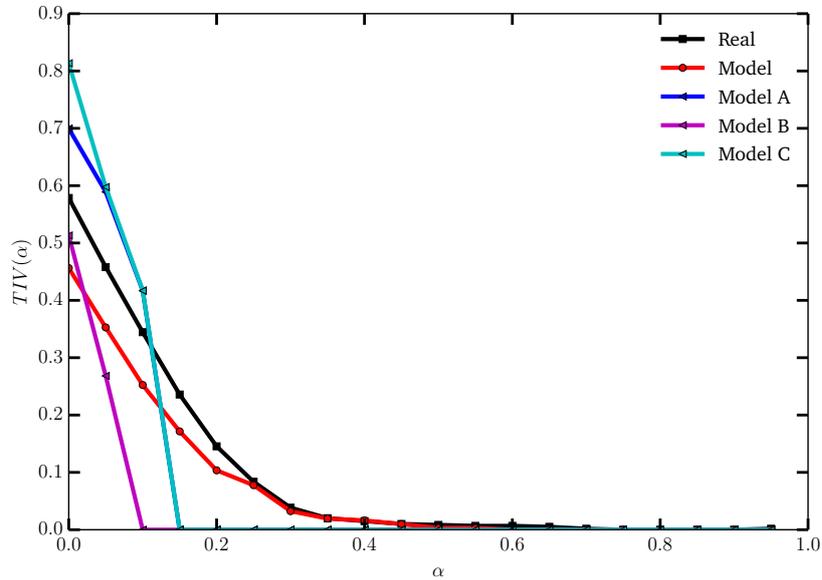

SUPPLEMENTARY FIGURE 25. **Triangle inequality violation spectrum.** Comparison between $TIV(\alpha)$ curves measured for the real US Commodities network, our model and the models A, B and C with the exponents given in the caption of Supplementary Figures 22–24.

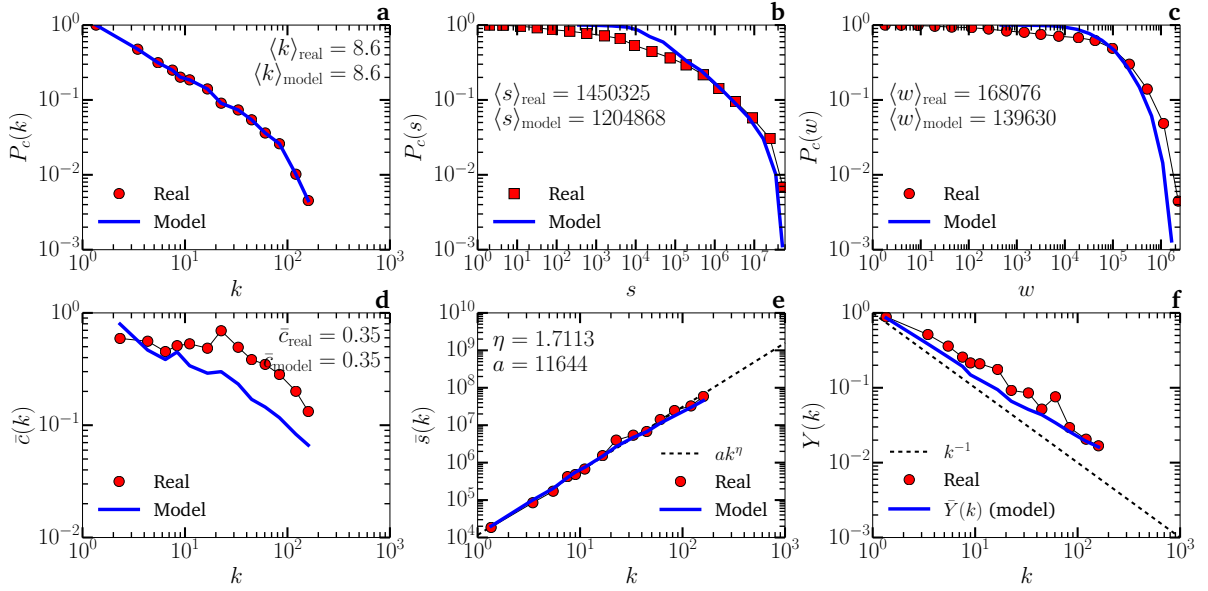

SUPPLEMENTARY FIGURE 26. **Predictions by Model A.** Comparison between topological and weighted properties of the US airports network (symbols) and a synthetic network generated by model A with $\theta = 0.89$ (solid lines). **a**, complementary cumulative degree distribution. **b**, complementary cumulative strength distribution. **c**, complementary cumulative weight distribution of links. **d**, degree-dependent clustering coefficient. **e**, average strength of nodes of degree $k$. **f**, disparity of nodes as a function of their degree.

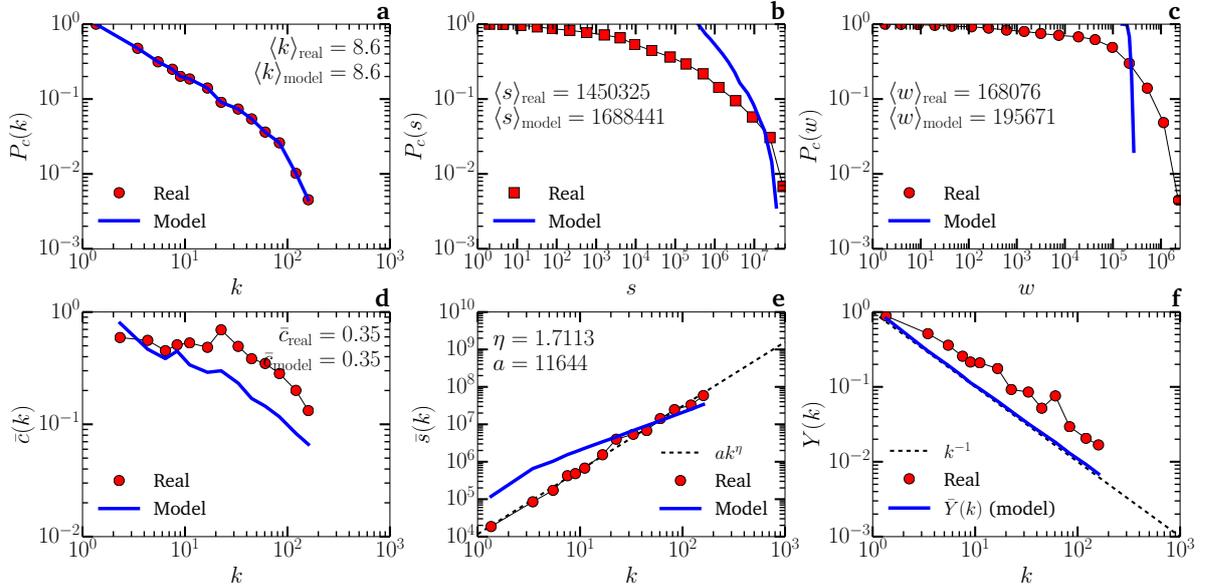

SUPPLEMENTARY FIGURE 27. **Predictions by Model B.** Comparison between topological and weighted properties of the US airports network (symbols) and a synthetic network generated by model B with $\delta = -0.12$ (solid lines). **a**, complementary cumulative degree distribution. **b**, complementary cumulative strength distribution. **c**, complementary cumulative weight distribution of links. **d**, degree-dependent clustering coefficient. **e**, average strength of nodes of degree $k$. **f**, disparity of nodes as a function of their degree.

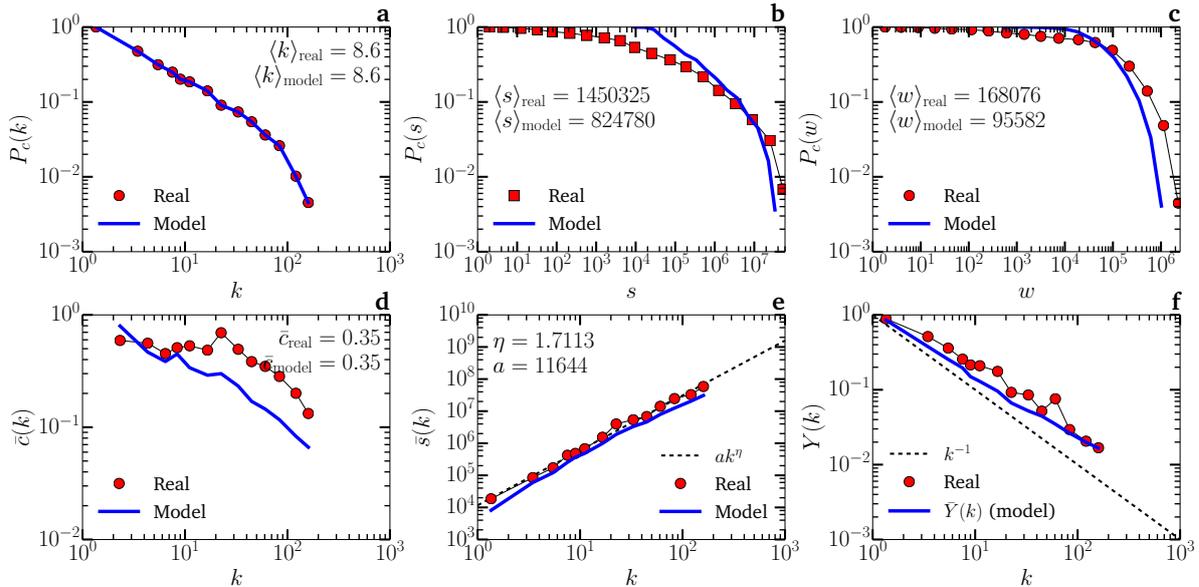

SUPPLEMENTARY FIGURE 28. **Predictions by Model C.** Comparison between topological and weighted properties of the US airports network (symbols) and a synthetic network generated by model C with $\mu = 1.05$ and $\nu = 0.225$ (solid lines). **a**, complementary cumulative degree distribution. **b**, complementary cumulative strength distribution. **c**, complementary cumulative weight distribution of links. **d**, degree-dependent clustering coefficient. **e**, average strength of nodes of degree $k$. **f**, disparity of nodes as a function of their degree.

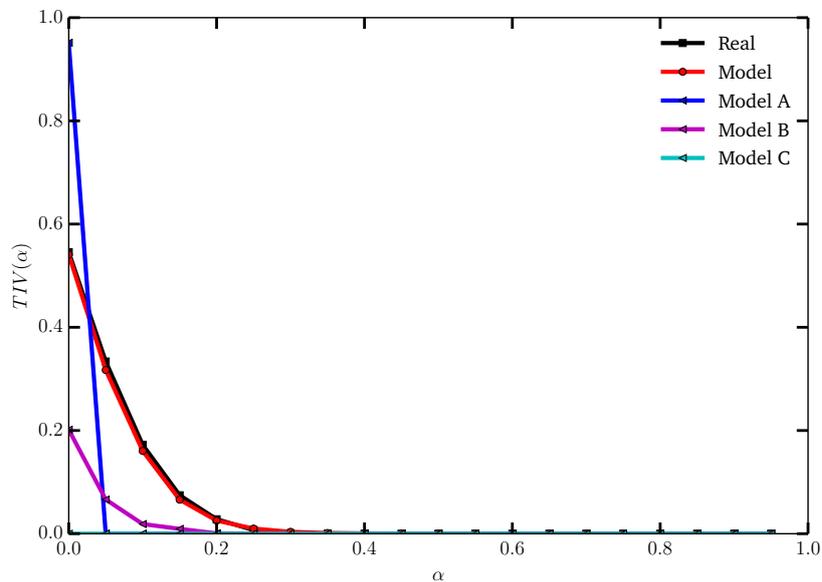

SUPPLEMENTARY FIGURE 29. **Triangle inequality violation spectrum.** Comparison between $TIV(\alpha)$ curves measured for the real US airports network, our model and the models A, B and C with the exponents given in the caption of Supplementary Figures 26–28.

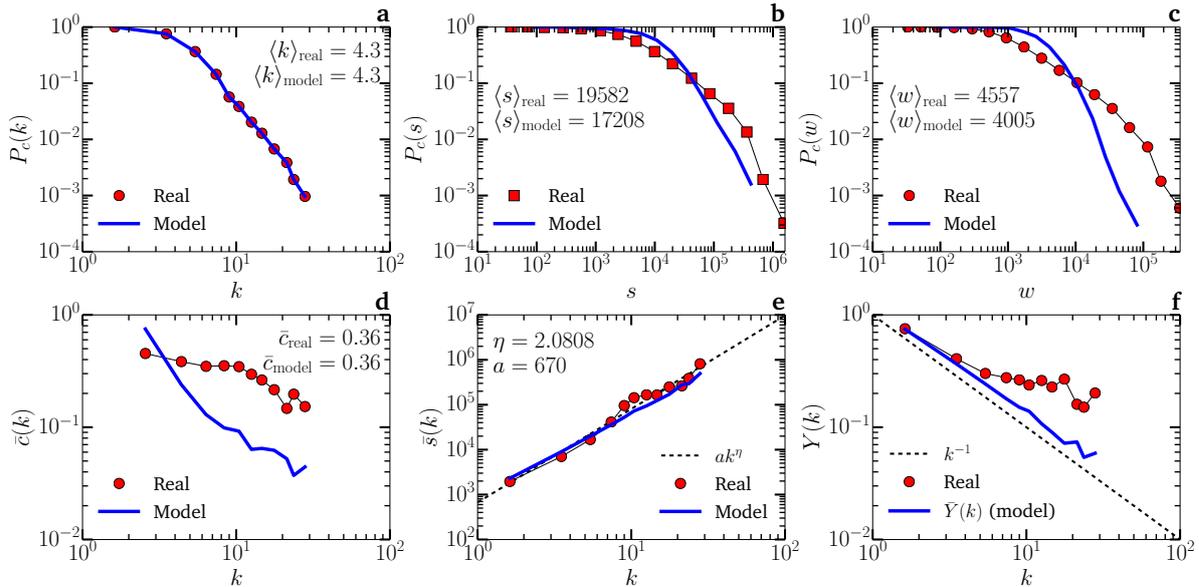

SUPPLEMENTARY FIGURE 30. **Predictions by Model A.** Comparison between topological and weighted properties of the US Commute network (symbols) and a synthetic network generated by model A with $\theta = 1.07$ (solid lines). **a**, complementary cumulative degree distribution. **b**, complementary cumulative strength distribution. **c**, complementary cumulative weight distribution of links. **d**, degree-dependent clustering coefficient. **e**, average strength of nodes of degree $k$. **f**, disparity of nodes as a function of their degree.

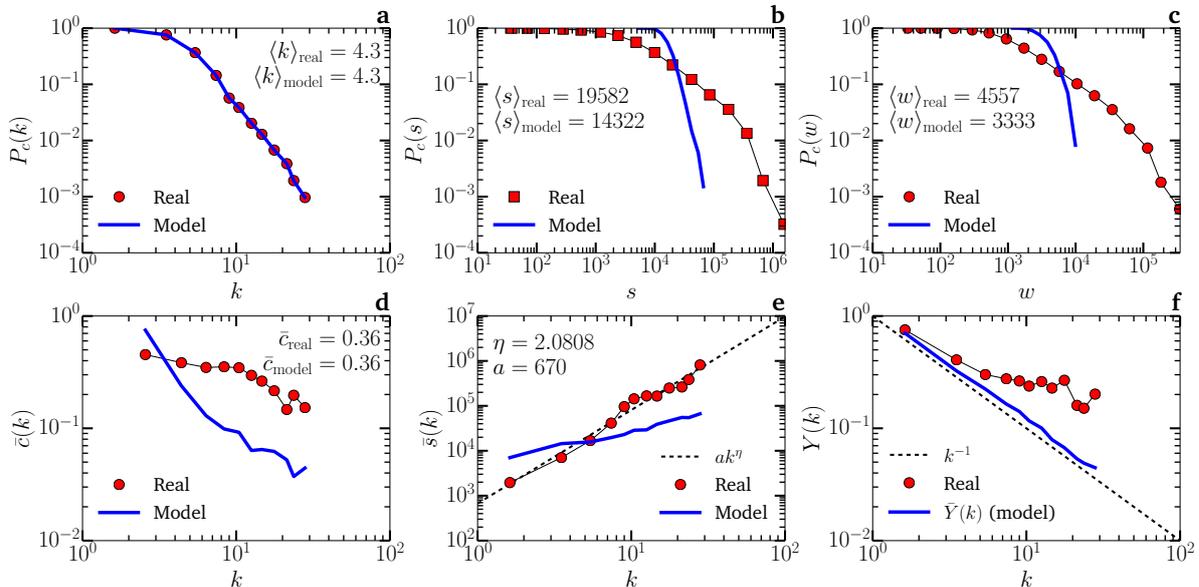

SUPPLEMENTARY FIGURE 31. **Predictions by Model B.** Comparison between topological and weighted properties of the US Commute network (symbols) and a synthetic network generated by model B with $\delta = 0.32$ (solid lines). **a**, complementary cumulative degree distribution. **b**, complementary cumulative strength distribution. **c**, complementary cumulative weight distribution of links. **d**, degree-dependent clustering coefficient. **e**, average strength of nodes of degree $k$. **f**, disparity of nodes as a function of their degree.

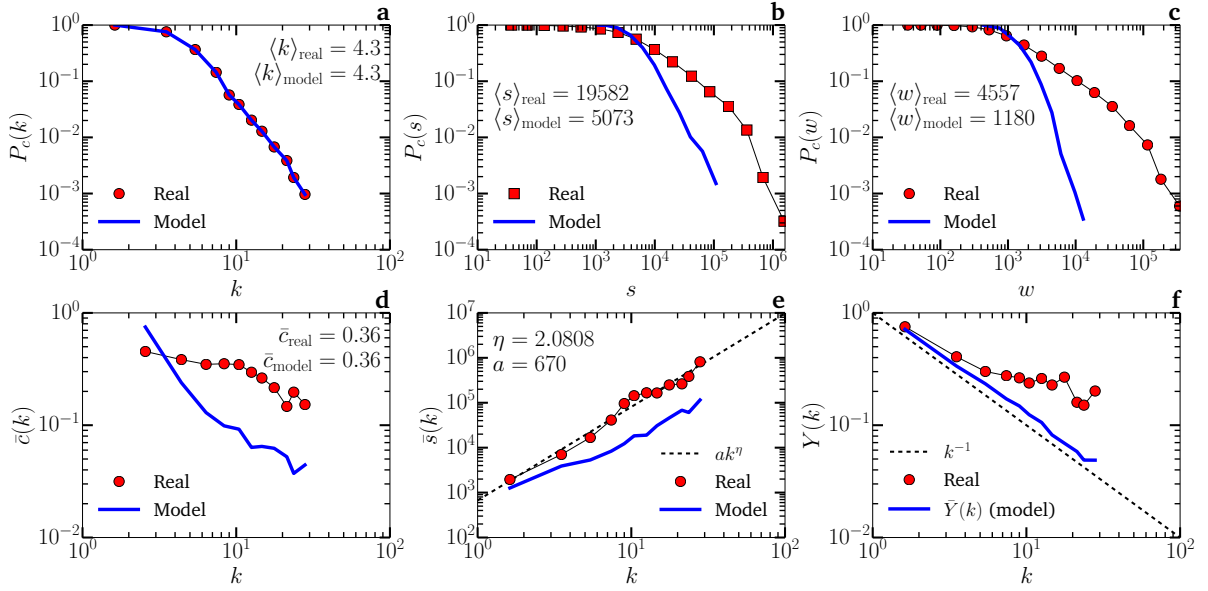

SUPPLEMENTARY FIGURE 32. **Predictions by Model C.** Comparison between topological and weighted properties of the US Commute network (symbols) and a synthetic network generated by model C with $\mu = 1.35$ and $\nu = 0.625$ (solid lines). **a**, complementary cumulative degree distribution. **b**, complementary cumulative strength distribution. **c**, complementary cumulative weight distribution of links. **d**, degree-dependent clustering coefficient. **e**, average strength of nodes of degree $k$. **f**, disparity of nodes as a function of their degree.

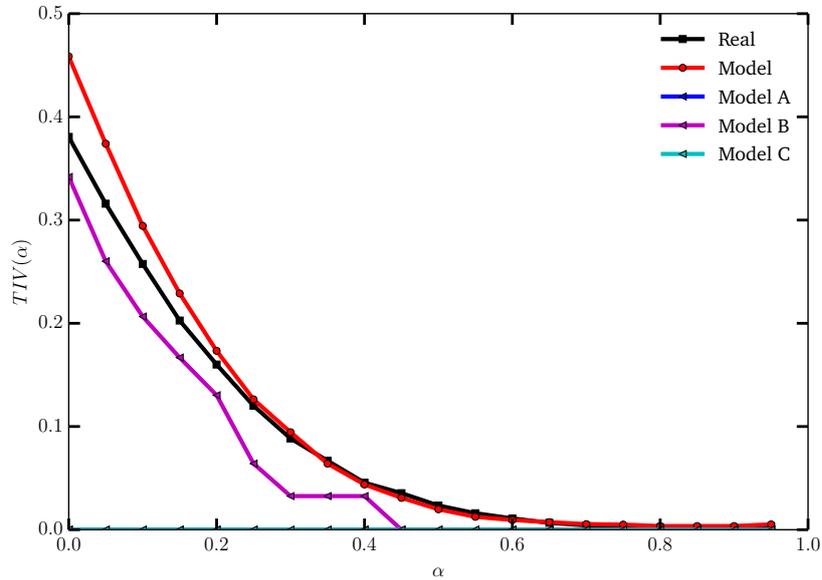

SUPPLEMENTARY FIGURE 33. **Triangle inequality violation spectrum.** Comparison between $TIV(\alpha)$ curves measured for the real US Commute network, our model and the models A, B and C with the exponents given in the caption of Supplementary Figures 30–32.

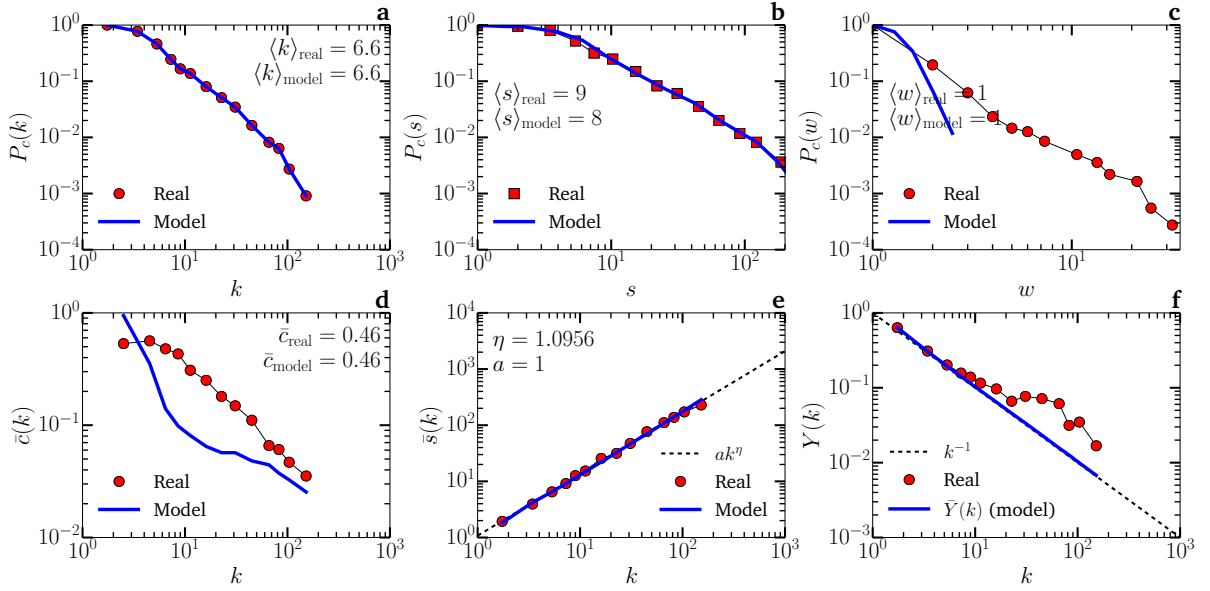

SUPPLEMENTARY FIGURE 34. **Predictions by Model A.** Comparison between topological and weighted properties of the iJO1366 *E. Coli* metabolic network (symbols) and a synthetic network generated by model A with $\theta = 0.16$ (solid lines). **a,** complementary cumulative degree distribution. **b,** complementary cumulative strength distribution. **c,** complementary cumulative weight distribution of links. **d,** degree-dependent clustering coefficient. **e,** average strength of nodes of degree $k$. **f,** disparity of nodes as a function of their degree.

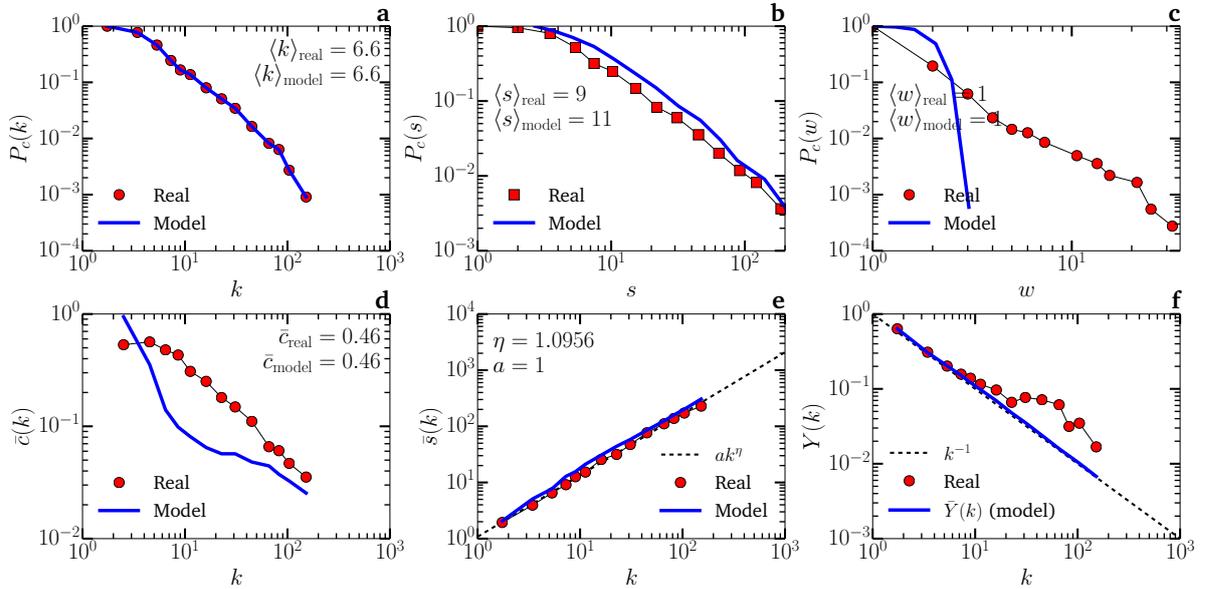

SUPPLEMENTARY FIGURE 35. **Predictions by Model B.** Comparison between topological and weighted properties of the iJO1366 *E. Coli* metabolic network (symbols) and a synthetic network generated by model B with $\delta = -0.15$ (solid lines). **a,** complementary cumulative degree distribution. **b,** complementary cumulative strength distribution. **c,** complementary cumulative weight distribution of links. **d,** degree-dependent clustering coefficient. **e,** average strength of nodes of degree $k$. **f,** disparity of nodes as a function of their degree.

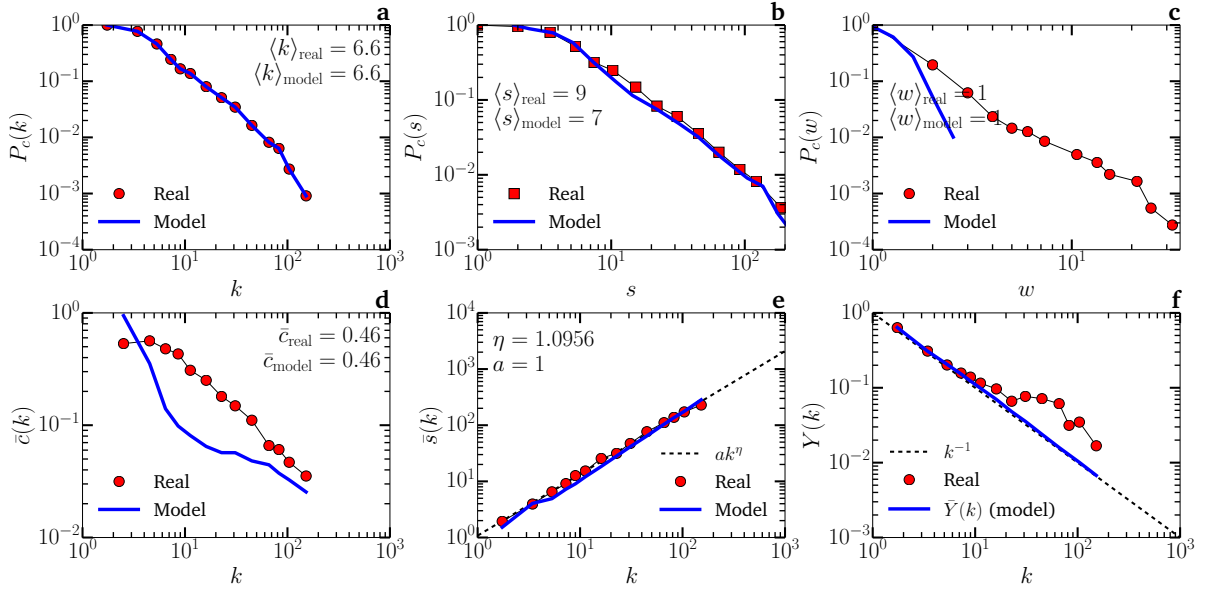

SUPPLEMENTARY FIGURE 36. **Predictions by Model C.** Comparison between topological and weighted properties of the iJO1366 *E. Coli* metabolic network (symbols) and a synthetic network generated by model C with $\mu = 0.225$ and $\nu = 0.075$ (solid lines). **a**, complementary cumulative degree distribution. **b**, complementary cumulative strength distribution. **c**, complementary cumulative weight distribution of links. **d**, degree-dependent clustering coefficient. **e**, average strength of nodes of degree $k$. **f**, disparity of nodes as a function of their degree.

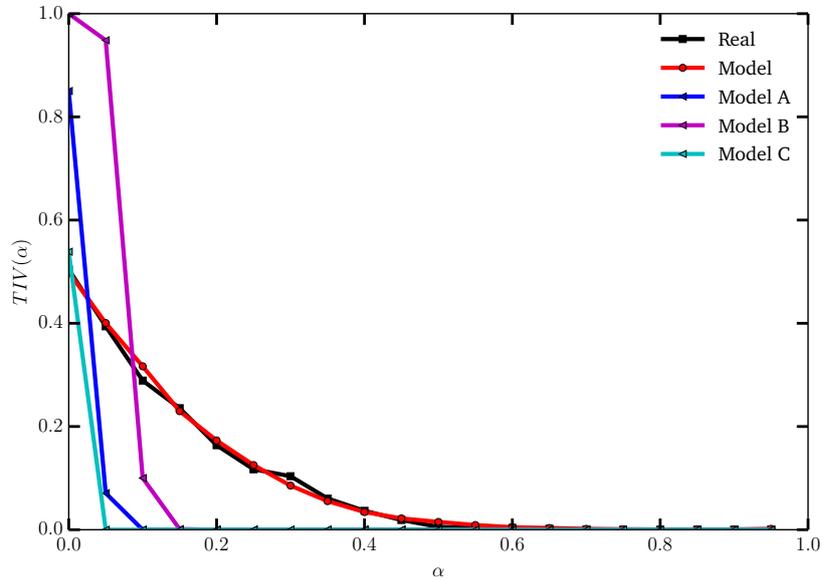

SUPPLEMENTARY FIGURE 37. **Triangle inequality violation spectrum.** Comparison between $TIV(\alpha)$ curves measured for the real iJO1366 *E. Coli* metabolic network, our model and the models A, B and C with the exponents given in the caption of Supplementary Figures 34–36.

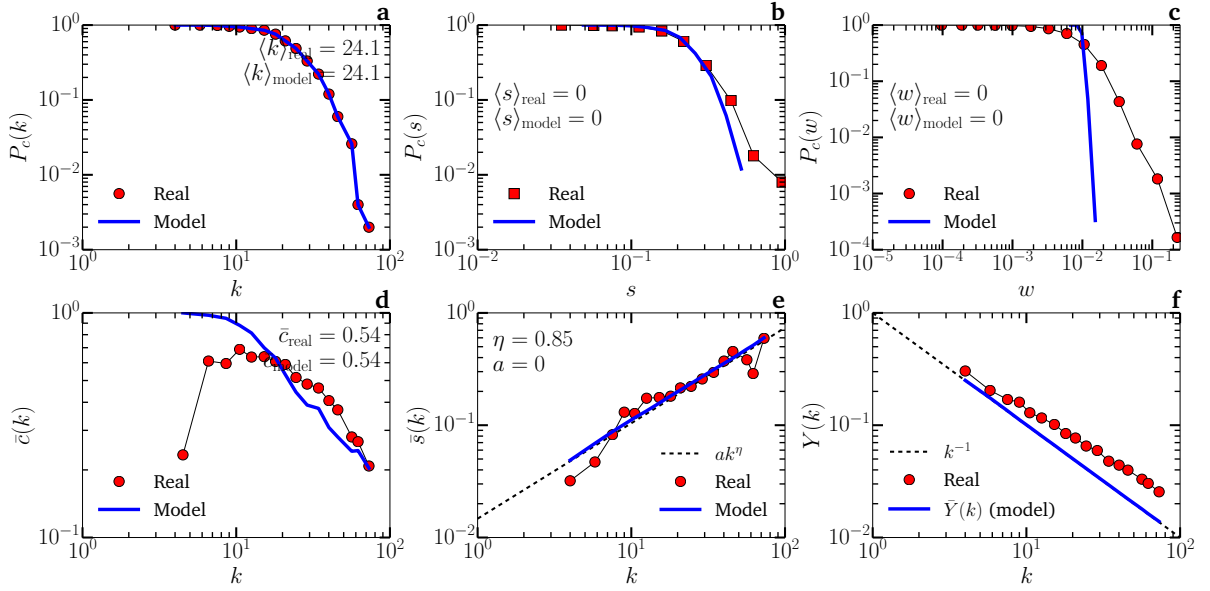

SUPPLEMENTARY FIGURE 38. **Predictions by Model A.** Comparison between topological and weighted properties of the Human brain network (symbols) and a synthetic network generated by model A with $\theta = -0.19$ (solid lines). **a**, complementary cumulative degree distribution. **b**, complementary cumulative strength distribution. **c**, complementary cumulative weight distribution of links. **d**, degree-dependent clustering coefficient. **e**, average strength of nodes of degree $k$. **f**, disparity of nodes as a function of their degree.

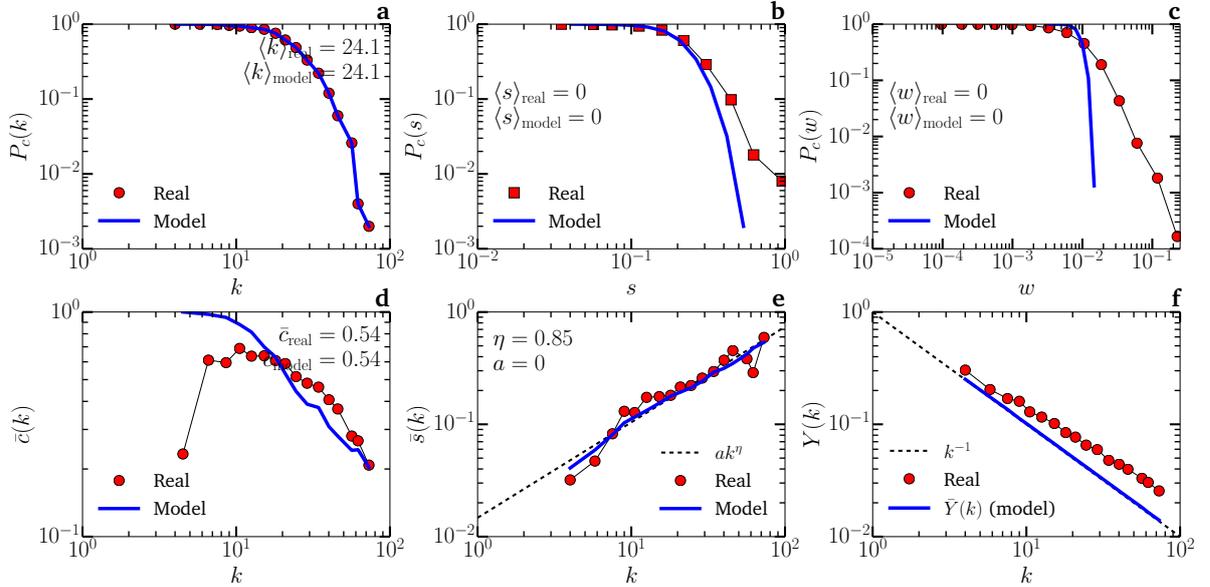

SUPPLEMENTARY FIGURE 39. **Predictions by Model B.** Comparison between topological and weighted properties of the Human brain network (symbols) and a synthetic network generated by model B with $\delta = 0.33$ (solid lines). **a**, complementary cumulative degree distribution. **b**, complementary cumulative strength distribution. **c**, complementary cumulative weight distribution of links. **d**, degree-dependent clustering coefficient. **e**, average strength of nodes of degree $k$. **f**, disparity of nodes as a function of their degree.

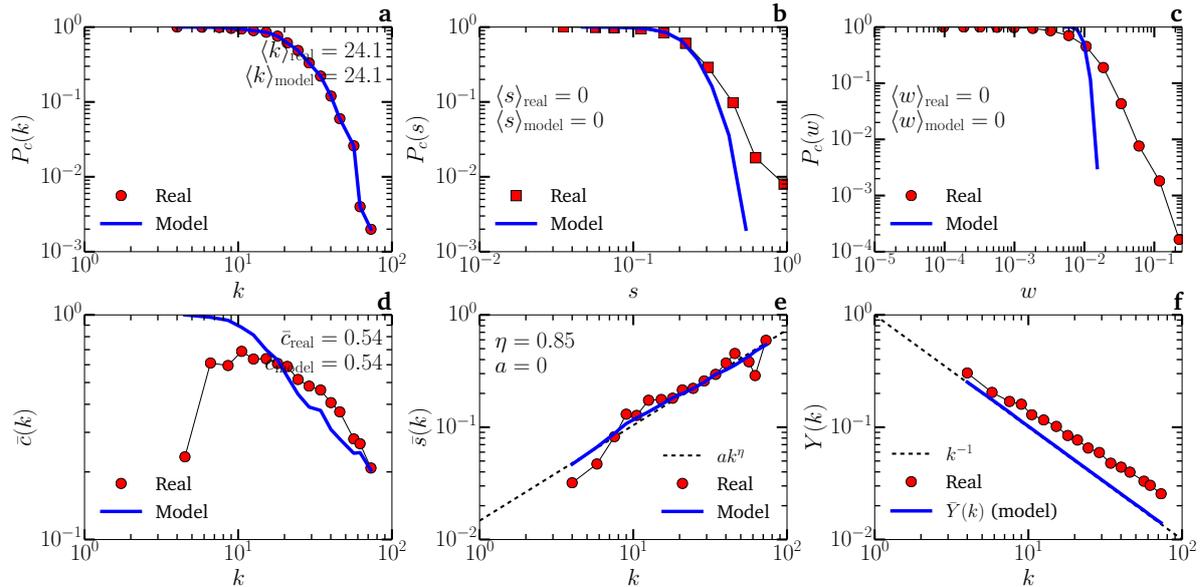

SUPPLEMENTARY FIGURE 40. **Predictions by Model C.** Comparison between topological and weighted properties of the Human brain network (symbols) and a synthetic network generated by model C with $\mu = -0.125$ and $\nu = 0.2$ (solid lines). **a**, complementary cumulative degree distribution. **b**, complementary cumulative strength distribution. **c**, complementary cumulative weight distribution of links. **d**, degree-dependent clustering coefficient. **e**, average strength of nodes of degree $k$. **f**, disparity of nodes as a function of their degree.

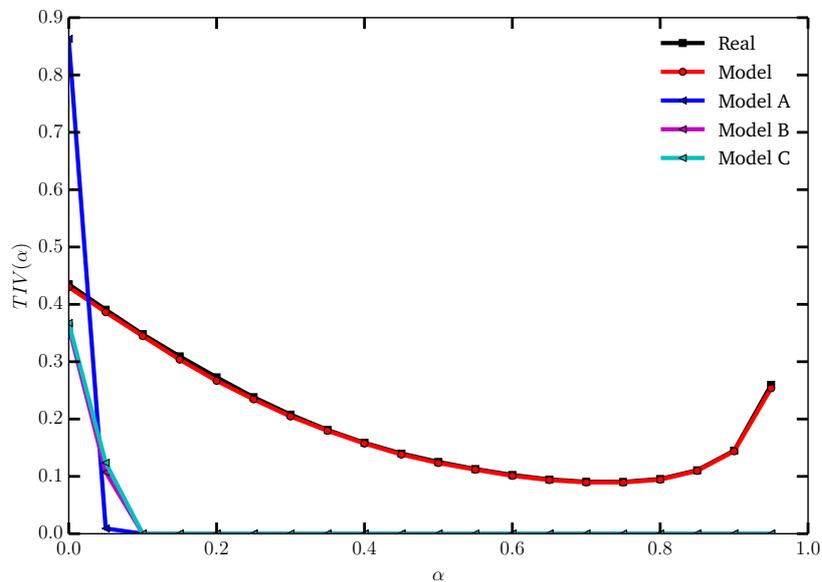

SUPPLEMENTARY FIGURE 41. **Triangle inequality violation spectrum.** Comparison between $TIV(\alpha)$ curves measured for the real Human brain network, our model and the models A, B and C with the exponents given in the caption of Supplementary Figures 38–40.

# Supplementary References


\end{document}